\documentclass[journal]{IEEEtran}

\newcommand{\HISTW}{0.56} 
\newcommand{\ARW}{0.60} 

\usepackage{graphicx,xcolor}
\usepackage[OT1]{fontenc}
\usepackage[numbers,sort&compress]{natbib}
\usepackage[cmex10]{amsmath}
\usepackage{amssymb}
\usepackage{bm}
\usepackage{graphicx}
\usepackage{subfigure}
\usepackage{tabularx}
\usepackage{arydshln}
\usepackage{mathtools}
\usepackage{multirow}
\usepackage{algorithm,algorithmic}
\usepackage{url}
\usepackage{listings}
\usepackage{comment}
\usepackage{textcase}
\usepackage{hyperref}
\usepackage[absolute,overlay]{textpos}

\def\E{\mathbf{E}}
\def\cX{\mathcal{X}}
\def\cS{\mathcal{S}}
\def\D{\mathbf{D}}
\def\I{\mathbf{I}}
\def\Y{\mathbf{Y}}

\def\H{\mathbf{H}}
\def\Hh{\hat{\H}}

\def\Hb{\bar{\H}}
\def\Hrm{\mathrm{H}}

\def\sTM{\sqrt{\frac{T}{M}}}
\def\TM{\frac{T}{M}}
\def\Trm{\mathrm{T}}
\def\MT{\frac{M}{T}}
\def\tr{\mathrm{tr}}

\def\V{\mathbf{V}}
\def\Vp{\mathbf{V}^{\prime}}

\def\x{\mathbf{x}}
\def\X{\mathbf{X}}

\def\S{\mathbf{S}}

\def\Frm{\mathrm{F}}

\newcommand{\maximize}{\mathop{\rm maximize}\limits}
\newcommand{\minimize}{\mathop{\rm minimize}\limits}

\makeatletter
\def\ps@IEEEtitlepagestyle{%
  \def\@oddfoot{\mycopyrightnotice}%
  \def\@oddhead{\hbox{}\@IEEEheaderstyle\leftmark\hfil\thepage}\relax
  \def\@evenhead{\@IEEEheaderstyle\thepage\hfil\leftmark\hbox{}}\relax
  \def\@evenfoot{}%
}
\def\mycopyrightnotice{%
  \begin{minipage}{\textwidth}
  \centering \scriptsize
\textcopyright 2024 IEEE. Personal use of this material is permitted.
  Permission from IEEE must be obtained for all other uses, in any current or future
  media, including reprinting/republishing this material for advertising or promotional
  purposes, creating new collective works, for resale or redistribution to servers or
  lists, or reuse of any copyrighted component of this work in other works.
  DOI: \href{https://doi.org/10.1109/TWC.2024.3369091}{10.1109/TWC.2024.3369091}
  \end{minipage}
}
\makeatother
\begin{document}
\title{Boosting Spectral Efficiency with Data-Carrying Reference Signals on the Grassmann Manifold}

\author{
Naoki~Endo,~\IEEEmembership{Graduate Student Member,~IEEE},
Hiroki~Iimori,~\IEEEmembership{Member,~IEEE},\\
Chandan~Pradhan,~\IEEEmembership{Member,~IEEE},
Szabolcs~Malomsoky,
and Naoki~Ishikawa,~\IEEEmembership{Senior Member,~IEEE}.
\thanks{N.~Endo and N.~Ishikawa are with the Faculty of Engineering, Yokohama National University, 240-8501 Kanagawa, Japan (e-mail: ishikawa-naoki-fr@ynu.ac.jp). H.~Iimori, C.~Pradhan, and S.~Malomsoky are with Ericsson Research, Ericsson Japan K.K., 220-0012 Kanagawa, Japan.
}}

\markboth{\NoCaseChange{March 1, 2024}}
{Shell \MakeLowercase{\textit{et al.}}: Bare Demo of IEEEtran.cls for Journals}
\maketitle

\TPshowboxesfalse
\begin{textblock*}{\textwidth}(45pt,10pt)
\footnotesize
\centering
Accepted for publication in IEEE Transactions on Wireless Communications. This is the author's version which has not been fully edited and content may change prior to final publication. Citation information: DOI 10.1109/TWC.2024.3369091
\end{textblock*}

\begin{abstract}
In wireless networks, frequent reference signal transmission for accurate channel reconstruction may reduce spectral efficiency. To address this issue, we consider to use a data-carrying reference signal (DC-RS) that can simultaneously estimate channel coefficients and transmit data symbols. Here, symbols on the Grassmann manifold are exploited to carry additional data and to assist in channel estimation. Unlike conventional studies, we analyze the channel estimation errors induced by DC-RS and propose an optimization method that improves the channel estimation accuracy without performance penalty. Then, we derive the achievable rate of noncoherent Grassmann constellation assuming discrete inputs in multi-antenna scenarios, as well as that of coherent signaling assuming channel estimation errors modeled by the Gauss-Markov uncertainty. These derivations enable performance evaluation when introducing DC-RS, and suggest excellent potential for boosting spectral efficiency, where interesting crossings with the non-data carrying RS occurred at intermediate signal-to-noise ratios.
\end{abstract}


\IEEEpeerreviewmaketitle

\section{Introduction\label{sec:intro}}
Reference signals (RS) enable estimating the channel state information (CSI) between a transmitter and a receiver.
The estimated CSI plays an important role in resource allocation, adaptive modulation, beam control, and signal demodulation.
For example, 5G NR defines CSI-RS, demodulation RS (DM-RS), phase tracking RS (PT-RS), and sounding RS (SRS) \cite{5gnrch} for different channel estimation purposes.

The overhead of pilot symbols (such as DM-RS in 5G) is adjusted in an adaptive manner depending on the communication environment, reaching a maximum transmission ratio of approximately 30\% \cite{5gnrch}.
Pilot symbols are known a priori between the transmitter and the receiver, and a channel estimation method using such RS is referred to as a training method.
The training method has low computational complexity and high estimation accuracy, therefore it is widely used in wireless networks, even though the cost of high overhead can arise in certain scenarios, such as, for example, high-mobility scenarios due to frequent CSI updates.

Is it possible to reduce the channel estimation overhead to near zero?
There are a number of related studies \cite{hassibi2003how}, which can be categorized according to whether the presented method requires a preamble, which is a kind of RS.
One approach that requires a preamble is differential coding, which takes advantage of the relationship between symbols, such as a phase difference \cite{weber1978differential}, a rotation by unitary matrix \cite{tarokh2000differential}, a look-up table \cite{wei2011differential}, and a space-time projection \cite{ishikawa2018differential}.
The relationship allows for noncoherent detection and stable performance even in extreme high-mobility scenarios; however, it induces the well-known 3 dB performance loss due to doubled noise power \cite{weber1978differential,tarokh2000differential}.
The other approach is the superimposed pilot method \cite{verenzuela2018spectral,upadhya2018downlink}. It improves spectral efficiency (SE) by superimposing data and RS in the same radio resources, which induces inevitable mutual interference and error floors.
The semi-blind method \cite{medles2001semiblind,chen2010semiblind,iimori2021TWC} can be interpreted as a training method that intends to reduce the number of RSs as much as possible.

An approach that completely eliminates the need for preamble transmission is referred to as a blind method.
For example, the pioneering study by Shabazpanahi et al. \cite{shabazpanahi2005blind} takes advantage of the orthogonality of space-time block codes to estimate CSI from the second-order statistics of received signals.
However, the lack of preamble transmission introduces phase uncertainties in the CSI estimation, which must be resolved by relying on inefficient power allocation and asymmetric symbols.

It is desirable to remove the phase uncertainty while increasing the degree of freedom of symbol representation, in order to maximize spectral efficiency (SE).
The only known method to achieve this is based on using a certain topological representation of multidimensional spaces such as the Grassmann manifold \cite{zheng2002communication,gohary2019noncoherent}.
Points on the Grassmann manifold are equivalent with respect to rotation and scaling, which is known as an equivalence relation.
That is, a Grassmann constellation does not change on the manifold even when multiplied by a channel matrix, allowing for noncoherent detection \cite{zheng2002communication}.
The construction methods for Grassmann constellations can be classified into three main approaches: mapping of classic symbols \cite{kammoun2003new,kammoun2007noncoherent}, algebraic construction \cite{ngo2020cubesplit,konishi2022novel}, and numerical optimization \cite{gohary2009noncoherent,elmossallamy2019noncoherent}.
The method relying on numerical optimization is optimal in terms of reliability and achievable rate, while the algebraic construction is optimal in terms of detection complexity due to its systematic structure \cite{ngo2020cubesplit}.

The idea of using the Grassmann constellation as an RS is not new.
Dating back to 2002, Warrier and Madhow claimed in \cite{warrier2002spectrally} that the noncoherent detection can be regarded as a joint estimation of channel and data symbols.
Similarly, Kammoun et al. considered the noncoherent detection as a closest point search over the coherent code, where the channel matrix was obtained from a noncoherently detected Grassmann codeword \cite{kammoun2007noncoherent}.
A more explicit idea was proposed by Yu et al. in \cite{yu2007informationbearing} that noncoherent codewords were used as \textit{data-carrying RS (DC-RS)}\footnote{In \cite{yu2007informationbearing}, the original concept was termed \textit{information-bearing noncoherently modulated pilots}. Due to the similarity with DM-RS, we term it as DC-RS in this paper, which has the same meaning.}, and the asymptotic signal-to-noise ratio (SNR) loss induced by DC-RS was analyzed in an information-theoretic manner.
Later, the idea of DC-RS was applied to turbo-coded scenarios \cite{chen2011turbo}.
From another perspective, a scheme that simultaneously transmits two types of codewords \cite{hussien2014multiresolution,seddik2017multiresolution,elmossallamy2019multiresolution}, Grassmann manifold and unitary group, can be considered a similar study because it allows for both coherent and noncoherent detection.

Given this background, we consider the use of DC-RS \cite{yu2007informationbearing} on the Grassmann manifold that simultaneously estimates data symbols and channel coefficients.
Despite the noncoherent detection, the estimated Grassmann codeword can be used for accurate channel estimation for subsequent high-rate data symbols, circumventing the issue of phase uncertainty.
The major contributions are summarized as follows.
\begin{enumerate}
\item We devise a method to improve the channel estimation accuracy of DC-RS. The equivalence relation on the Grassmann manifold is exploited to maximize channel estimation accuracy without performance penalty. This
perspective has not been considered in the conventional
studies \cite{warrier2002spectrally,kammoun2007noncoherent,yu2007informationbearing,chen2011turbo,seddik2017multiresolution,elmossallamy2019multiresolution}.
\item We derive the achievable rate of noncoherent Grassmann constellations assuming discrete inputs in multi-antenna scenarios as well as that of coherent signaling assuming CSI errors. Our analysis clarifies that enhancing channel estimation accuracy is equivalent to maximizing SE, and DC-RS is capable of boosting SE when applied as a pilot instead of a more traditional non-data carrying pilot, where interesting crossings with the classical training method occur at intermediate SNRs.
\end{enumerate}

The remainder of this paper is organized as follows. Conventional channel estimation methods are reviewed in Section~\ref{sec:chan}, and conventional construction methods for Grassmann constellations are reviewed in Section~\ref{sec:conv}.
Theoretical analysis of channel estimation accuracy is presented in Section~\ref{sec:prop}, and the achievable rates of noncoherent and coherent signaling are derived in Section~\ref{sec:ana}.
Then, the performance of DC-RS is evaluated in Section~\ref{sec:performance}.
Finally, we conclude this paper in Section~\ref{sec:conc}.

\section{System Model and Conventional Channel Estimation Methods\label{sec:chan}}
The training method, which relies on RS for channel estimation, and the semi-blind method, which intends to reduce RS as much as possible, are reviewed in this section.
Note that the method using superimposed pilots is omitted because of the issue of error floors, and the blind estimation method is omitted because of the issue of phase uncertainty.

\subsection{Common System Model}
Let us consider a multiple-input multiple-output (MIMO) system with $M$ transmit antennas and $N$ receive antennas. Let $T ~ (\geq M)$ be the number of time slots for a space-time block code (STBC) $\S \in \mathbb{C}^{T \times M}$, which satisfies the power constraint of $\mathrm{E}[\| \S \|_{\Frm}^2] = T$.
A channel matrix $\H \in \mathbb{C}^{M \times N}$ is assumed to follow quasi-static Rayleigh fading and be invariant during $T$ slots. Let $\V \in \mathbb{C}^{T \times N}$ be additive white Gaussian noise, and each element of $\H$ and $\V$ follows $\mathcal{CN}(0,1)$.
The received symbol $\Y \in \mathbb{C}^{T \times N}$ is represented as
\begin{align}
    \Y = \S \H + \sigma_v \V, \label{eq:sys}
\end{align}
where the SNR is defined as $\mathrm{SNR} = 10 \cdot \mathrm{log}_{10} \left( 1 / \sigma_v^2 \right)$ [dB].

\subsection{Training Method\label{subsec:train}}
The training method that relies on RS is a standard in wireless communications since its complexity and performance are the best among representative channel estimation methods.
An RS matrix $\mathbf{P} \in \mathbb{C}^{T \times M}$ with $T \geq M$ is assumed to be known at the transmitter and receiver.
As a concrete example, DM-RS, a type of RS in the 5G NR system, is a pseudo-random sequence drawn from a set \cite{5gnrch}
\begin{align}
\left\{
\frac{1}{\sqrt{2}} + \frac{j}{\sqrt{2}},
\frac{1}{\sqrt{2}} - \frac{j}{\sqrt{2}},
-\frac{1}{\sqrt{2}} + \frac{j}{\sqrt{2}},
-\frac{1}{\sqrt{2}} - \frac{j}{\sqrt{2}}
\right\},
\end{align}
which are quadrature phase-shift keying (QPSK) symbols.
According to the system model \eqref{eq:sys}, the received symbol is represented as $\Y = \mathbf{P}\H + \sigma_v \V$.
Here, the estimate of $\H$, denoted by $\Hh$, is simply given by the zero-forcing (ZF) or minimum mean square error (MMSE) equalization of
\begin{align}
\Hh = \begin{cases}
    \mathbf{P}^{+} \Y & \textrm{(ZF)} \\
    (\mathbf{P}^{\Hrm} \mathbf{P} + \sigma^2_v\I_M)^{-1} \mathbf{P}^{\Hrm} \Y & \textrm{(MMSE)} \\
\end{cases}
\label{eq:HhatZFMMSE}
\end{align}
where we have a pseudo-inverse matrix of
\begin{align}
\mathbf{P}^{+} =
(\mathbf{P}^{\Hrm} \mathbf{P})^{-1}\mathbf{P}^{\Hrm}
\end{align}
for $T \geq M$.

It is worth noting that depending on the properties of $\mathbf{P}$, such as $\mathbf{P}$ being an identity or a tall unitary matrix satisfying $\mathbf{P}^{\Hrm} \mathbf{P} = \I_M$, the inverse matrix calculation in \eqref{eq:HhatZFMMSE} might be simplified. This channel estimation process requires the lowest computational complexity among the counterparts mentioned earlier.
Specifically, in the ZF case, the computational complexity represented by the number of real multiplications is $\mathcal{O}(TMN)$, while in the MMSE case, it is $\mathcal{O}(M^3)$.
However, frequent insertion of RSs is essential for high-mobility scenarios to update $\Hh$, resulting in SE degradation.

\subsection{Semi-Blind Method \cite{medles2001semiblind,chen2010semiblind}}
The semi-blind method is an extension of the training method that aims at reducing the number of RSs as much as possible, where the estimations of the channel matrix $\H$ and the codeword $\S$ are repeated in an iterative manner \cite{chen2010semiblind}.

Let $i$ be a transmission index, and the system model given in \eqref{eq:sys} is modified to $\Y(i) =  \S(i) \H + \sigma_v \V(i)$.
At the beginning, $i = 0$, $\S(0) = \I_M$ is transmitted.
In the first trial, a rough estimate of $\H$ is obtained by $\Hh = \Y(0)$.
For the following $W$ blocks, the semi-blind method first estimates codewords $\bar{\S} = [\hat{\S}(1)^{\Trm} ~ \cdots ~ \hat{\S}(W)^{\Trm}]^{\Trm} \in \mathbb{C}^{WT \times M}$.
Then, using $\bar{\S}$, the channel estimate is obtained by the ZF equalization of \cite{chen2010semiblind}
\begin{align}
\Hh = \bar{\mathbf{S}}^{+} \bar{\Y},
\end{align}
where we have concatenated received symbols as $\bar{\Y} = [\Y(1)^{\Trm} ~ \cdots ~ \Y(W)^{\Trm}]^{\Trm} \in \mathbb{C}^{WT \times N}$.
Using the refined $\Hh$, the estimation of $\bar{\S}$ becomes improved.
It is found in \cite{chen2010semiblind} that a few trials are sufficient to achieve the performance close to perfect CSI (PCSI) assumption.
Then, in our comparisons, we consider the PCSI case as a performance baseline, and omit the semi-blind method.

\section{Noncoherent Communication on the Grassmann Manifold and Classic DC-RS\label{sec:conv}}
This section summarizes the overview of the Grassmann manifold that enables the noncoherent detection of a space-time codeword.
In particular, three representative construction methods are introduced.

\subsection{Grassmann Manifold \cite{edelman1998geometryb,bendokat2020grassmann}}
Unitary matrices form a group as they are closed under multiplication.
Let $M$ be a non-negative integer.
The unitary group $\mathcal{U}(M)$ is a set of $M \times M$ unitary matrices, i.e., \cite{edelman1998geometryb}\footnote{In \cite{edelman1998geometryb}, the definitions are given for real values, but extending them to complex values is a straightforward task \cite{kammoun2003new}.}
\begin{align}
\mathcal{U}(M) = \{\mathbf{U} \in \mathbb{C}^{M \times M} ~ | ~ \mathbf{U}^{\Hrm}\mathbf{U} = \mathbf{I}_M\}.
\end{align}

Next, let $T \geq M$ be a non-negative integer.
The Stiefel manifold $\mathcal{S}(T,M)$ is a set of $T$-by-$M$ ``tall-skinny'' unitary matrices, i.e., \cite{edelman1998geometryb}
\begin{align}
\mathcal{S}(T,M) = \{\mathbf{S} \in \mathbb{C}^{T\times M} ~ | ~ \mathbf{S}^{\Hrm}\mathbf{S} = \mathbf{I}_M \}.
\end{align}
We define an equivalence relation $\sim$ for $\mathbf{S}_1, ~ \mathbf{S}_2 \in \mathcal{S}(T, M)$ to be a relation such that $\mathrm{span}(\mathbf{S}_1) = \mathrm{span}(\mathbf{S}_2)$, where $\mathrm{span}(\mathbf{S})$ denotes a vector space spanned by the column vectors in $\mathbf{S}$.
Here, an equivalence class of $\mathbf{S} \in \mathcal{S}(T, M)$ is denoted by \cite{edelman1998geometryb}
\begin{align}
[\mathbf{S}] = \{\mathbf{S}_1 \in \mathcal{S}(T,M) ~ | ~ \mathbf{S}_1 \sim \mathbf{S}\}.
\label{eq:equivalentGr}
\end{align}

Finally, the Grassmann manifold is defined as a quotient space of the Stiefel manifold $\mathcal{S}(T, M)$ with respect to the equivalence relation, i.e., \cite{edelman1998geometryb}
\begin{align}
\mathcal{G}(T,M) = \mathcal{S}(T,M) ~ / ~ \mathcal{U}(M) = \{[\mathbf{S}] ~ | ~ \mathbf{S} \in \mathcal{S}(T,M)\}.
\label{eq:GTM}
\end{align}

That is, a point on the Grassmann manifold $\mathcal{G}(T, M)$ is invariant as long as a matrix is multiplied from the right.
Even a non-unitary matrix do not change the point on the Grassmann manifold since the original matrix and the multiplied matrix span the same subspace.
This mathematical property can be exploited to enable noncoherent detection.

\subsection{Conventional Grassmann Constellation \cite{kammoun2003new,kammoun2007noncoherent,ngo2020cubesplit,gohary2009noncoherent}}
We review three representative construction methods using the exponential map (exp-map) \cite{kammoun2003new,kammoun2007noncoherent}, an algebraic property \cite{ngo2020cubesplit}, and the manifold optimization (manopt) \cite{gohary2009noncoherent,ngo2020cubesplit}.

\subsubsection{Exp-Map \cite{kammoun2003new,kammoun2007noncoherent}}
A point $\X \in \mathbb{C}^{T \times M}$ on the Grassmann manifold $\mathcal{G}(T,M)$ can be represented by \cite{kammoun2007noncoherent}
\begin{align}
   \X = \left[ \mathrm{exp} \begin{pmatrix}
    \mathbf{0} & \mathbf{C} \\
    - \mathbf{C}^{\Hrm} & \mathbf{0}
    \end{pmatrix}
    \right] \mathbf{I}_{T,M}, \label{eq:exp-map}
\end{align}
where we have a complex-valued matrix
$\mathbf{C} \in \mathbb{C}^{M \times (T-M)}$, the matrix exponential function $\mathrm{exp}(\cdot)$, and $\mathbf{I}_{T,M} = [\mathbf{I}_{M \times M}\  \mathbf{0}_{M \times (T-M)}]^{\Trm}$.
The exp-map method proposed in \cite{kammoun2007noncoherent} uses the map of \eqref{eq:exp-map}.
Specifically, each element of $\mathbf{C} \in \mathbb{C}^{M\times(T-M)}$ is a quadrature amplitude modulation (QAM) symbol.
In the $M = 1$ case, $T-1$ symbols are simply mapped to each element of $\mathbf{C}$.
In the $M \geq 2$ case, the mapping rules are detailed in \cite{kammoun2007noncoherent}.
For example, if we consider $(M, T) = (2, 4)$, the matrix $\mathbf{C}$ is constructed by \cite{kammoun2007noncoherent}
\begin{align}
    \mathbf{C} = \begin{bmatrix}
        s_1 + \theta s_2 & \phi (s_3 + \theta s_4) \\
        \phi(s_3 - \theta s_4) & s_1 - \theta s_2
    \end{bmatrix},\label{eq:exp-map-c}
\end{align}
where $s_1, \cdots, s_4$ are QAM symbols and the parameters are $\phi^2 = \theta = e^{\mathrm{j} \pi / 4}$.

Let $L$ be a modulation order of QAM and $\cX$ be a codebook of Grassmann constellation.
The cardinality of exp-map is calculated as $|\cX| = L^{M (T-M)}$, and the transmission rate is
\begin{align}
R = \frac{\log_2 |\cX|}{T} = M \left(1 - \frac{M}{T} \right) \log_2(L) ~ \mathrm{[bit/sym]}.
\end{align}

In \cite{kammoun2007noncoherent}, a method is proposed to scale the QAM symbols and to maximize the minimum chordal distance (MCD) between codewords.
Maximizing MCD is equivalent to minimizing the upper bound of the symbol error ratio (SER), which is typically valid at high SNRs.
The complication here is that minimizing bit error ratio (BER) is not equivalent to minimizing SER and requires an appropriate bit labeling.
Since the QAM symbols are Gray coded, we can infer that the bit labeling after the map will be mostly better, but it is not clear whether the labeling is the best.
In general, Gray coding for a Grassmann constellation is known as an NP-hard problem \cite{colman2011quasigray}.

\subsubsection{Cube-Split \cite{ngo2020cubesplit}}
Cube-split is a method to construct a Grassmann constellation on $\mathcal{G}(T,1)$ algebraically without time-consuming optimization and is considered state-of-the-art at the time of writing.
Although the cube-split constellation is limited to single-antenna scenarios, i.e., $M=1$, it is approximately uniformly distributed on $\mathcal{G}(T,1)$ \cite{ngo2020cubesplit}, resulting in a higher MCD than the exp-map constellation.
The cardinality of cube-split is calculated as $|\cX| = T \cdot 2^{\sum_{j=1}^{2(T-1)} B_j}$, and the transmission rate is
\begin{align}
R = \frac{\log_2 |\cX|}{T} = \frac{1}{T} \log_2(T) + \frac{1}{T} \sum^{2(T-1)}_{j=1} B_j  ~ \mathrm{[bit/sym]},\label{eq:cubeR}
\end{align}
where $B_j \in \mathbb{Z}$
is the number of bits per time slot assigned for a real or imaginary part, a parameter that determines the transmission rate.
The construction procedure can be summarized in the following five main steps \cite{ngo2020cubesplit} that lead to uniformity of the constellation.
\begin{enumerate}
\item Determine the number of bits $B_j$ for $1\leq j \leq 2(T-1)$ satisfying a given transmission rate $R$.
To maximize MCD, from \eqref{eq:cubeR}, an even distribution is recommended so that the integer values $B_j$ are close to $(RT - \log_2(T)) / 2 / (T-1)$.

\item For each $B_j$, construct a set $A_j$ of points that divide the open interval $(0,1)$
\begin{align}
    A_j = \left\{\frac{1}{2^{B_j+1}},\frac{3}{2^{B_j+1}},\cdots,\frac{2^{B_j+1}-1}{2^{B_j+1}}  \right\}
\end{align}
and $|A_j| = 2^{B_j}$.
\item Take a direct product of all $A_j$ and represent their points on the grid by a vector
    \begin{align}
    \mathbf{a} = [a_1~ \cdots~ a_{2(T-1)}]^{\mathrm{T}} \in \bigotimes^{2(T-1)}_{j=1}A_j.
    \end{align}

\item For $T=2$, define a function $\mathbf{\xi}_1(\mathbf{a})$ for $\mathbf{a} = [a_1\ a_2]^{\Trm}$ as
\begin{align}
    \mathbf{\xi}_1(\mathbf{a}) = \sqrt{\frac{1-\exp(-\frac{|w|^2}{2})}{1+\exp(-\frac{|w|^2}{2})}}\frac{w}{|w|} \label{eq:cube-xi1}
\end{align}
with $w = N^{-1}(a_1) + \mathrm{j} N^{-1}(a_2)$, where $N(\cdot)$ is the cumulative distribution function (CDF) of the real Gaussian distribution and $\mathrm{j}$ is the imaginary number.
For $T>2$, define a function $\mathbf{\xi}_{T-1}(\mathbf{a})$ as
    \begin{align}
    &\mathbf{\xi}_{T-1}(\mathbf{a})
    = \mathbf{\xi}_{T-1}([a_1 ~ a_2 ~ \cdots ~ a_{2T-3} ~ a_{2T-2}]^{\Trm}) \nonumber \\
    &= \left[\mathbf{\xi}_{1}([a_1\ a_2]^{\Trm}) ~ \cdots ~ \mathbf{\xi}_{1}([a_{2T-3}\ a_{2T-2}]^{\Trm})\right]^{\Trm}. \label{xi2}
    \end{align}
\item Let a vector $\mathbf{t} = [t_1\cdots t_{T-1}]^{\Trm} = \mathbf{\xi}_{T-1}(\mathbf{a})$.
The cube-split constellation $\{\mathbf{g}_1(\mathbf{a}), \cdots, \mathbf{g}_{T}(\mathbf{a}) \}$
is constructed as
\begin{align}
    \mathbf{g}_i(\mathbf{a}) = \frac{[
    \overbrace{t_1 ~ \cdots ~ t_{i-1}}^{\text{Removed~if~}i=1}
    ~ 1 ~ t_i ~ \cdots ~ t_{T-1}]^{\Trm}}{\sqrt{1+\sum^{T-1}_{j=1}|t_j|^2}} \in \mathbb{C}^T
\end{align}
for $\mathbf{a} \in \bigotimes^{2(T-1)}_{j=1}A_j$ and $i=1, ~ \cdots ,T$.
\end{enumerate}

Let us check a specific example of the cube-split constellation for $(M,T)=(1,2)$.
If we set a desired transmission rate of $R=1.5$, the number of bits can be calculated as $B_1 = B_2 = (RT - \log_2(T)) / 2 / (T-1) = 1$.
Then, we have $A_1 = A_2 = \{1/4, 3/4\}$ and
\begin{align}
    \mathbf{a} =
    \begin{bmatrix}a_1 \\ a_2\end{bmatrix}
    \in
    \left\{
    \begin{bmatrix}1/4 \\ 1/4\end{bmatrix},
    \begin{bmatrix}1/4 \\ 3/4\end{bmatrix},
    \begin{bmatrix}3/4 \\ 1/4\end{bmatrix},
    \begin{bmatrix}3/4 \\ 3/4\end{bmatrix}
    \right\}.
    \label{eq:cubeconsta}
\end{align}
Here, since we have $N^{-1}(1/4) = -0.6745$ and $N^{-1}(3/4) = +0.6745$, from \eqref{eq:cube-xi1}, the corresponding $\mathbf{\xi}_1(\mathbf{a}) = \mathbf{t} = [t_1]$ values for \eqref{eq:cubeconsta} are calculated as
\begin{align}
    t_1 \in
    \{
    &-0.3344-0.3344\mathrm{j},
    -0.3344+0.3344\mathrm{j},\nonumber\\
    &+0.3344-0.3344\mathrm{j},
    +0.3344+0.3344\mathrm{j}
    \}
\end{align}
respectively.
In this case, the denominator of $\mathbf{g}_i(\mathbf{a})$ becomes a constant value, $\sqrt{1+|t_1|^2} = 1.1062$, and let the two symbols in $\mathbf{g}_i(\mathbf{a})$ be $c_1 = 1 / 1.1062 = 0.9040$ and $c_2 = (-0.3344-0.3344\mathrm{j}) / 1.1062 = -0.3023-0.3023\mathrm{j}$.
For each member of \eqref{eq:cubeconsta}, a subset
\begin{align}
    \left\{
        \begin{bmatrix} c_1   \\ c_2\end{bmatrix},
        \begin{bmatrix} c_1   \\ c_2^*\end{bmatrix},
        \begin{bmatrix} c_1   \\ -c_2^*\end{bmatrix},
        \begin{bmatrix} c_1   \\ -c_2\end{bmatrix}
        \right\}
        \label{eq:cubeex1}
\end{align}
is generated by $\mathbf{g}_1(\mathbf{a})$ and another subset
\begin{align}
    \left\{
        \begin{bmatrix} c_2   \\ c_1\end{bmatrix},
        \begin{bmatrix} c_2^* \\ c_1\end{bmatrix},
        \begin{bmatrix}-c_2^* \\ c_1\end{bmatrix},
        \begin{bmatrix}-c_2   \\ c_1\end{bmatrix}
        \right\}
        \label{eq:cubeex2}
\end{align}
is generated by $\mathbf{g}_{2}(\mathbf{a})$.
Finally, the union set of \eqref{eq:cubeex1} and \eqref{eq:cubeex2} is used as the cube-split constellation for $(M,T) = (1, 2)$ and $R=1.5$.

\begin{figure}[tb]
	\centering
        \includegraphics[clip, scale=0.65]{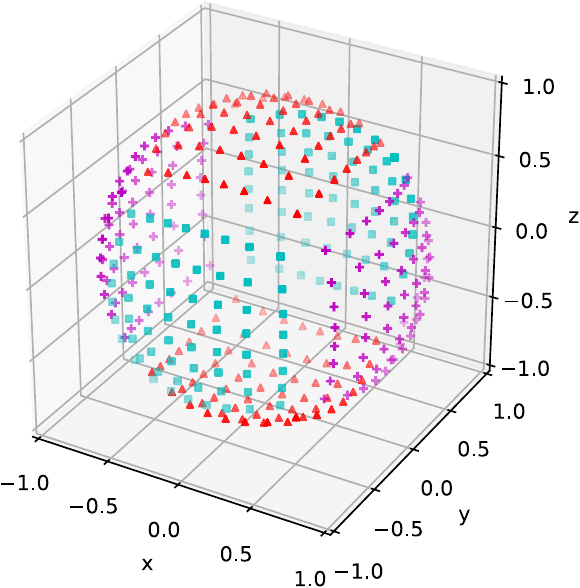}
	\caption{An example of cube-split \cite{ngo2020cubesplit} constellation.\label{fig:Cubesplit}}
\end{figure}
Furthermore, Fig.~\ref{fig:Cubesplit} shows the cube-split constellation having $(M, T)=(1, 3)$ and $(B_1, B_2) = (3, 3)$, which are determined for a simple 3D visualization, and no information is assigned to the imaginary part.
That is, the even-numbered elements of $\mathbf{a}$ is set to zero, which corresponds to the imaginary part of $\mathbf{t}$ being zero.
Since $A_1$ and $A_2$ are calculated as $\{1/16, 3/16, 5/16, 7/16, 9/16, 11/16, 13/16, 15/16\}$, $8 \cdot 8 = 64$ types of $\mathbf{a}$ are generated, such as $\mathbf{a} = [1/16 ~ 0 ~ 3/16 ~ 0]^{\Trm}$ and $[13/16 ~ 0 ~ 9/16 ~ 0]^{\Trm}$.
The cardinality here is $T \cdot 2^{B_1+B_2} = 3 \cdot 64 =192$.
Although $2 \cdot 192$ points are shown in Fig.~\ref{fig:Cubesplit}, all points symmetrical to the origin are considered to be the same point on the Grassmann manifold.
That is, any point multiplied by a scalar represents the same point on the manifold, and these mathematical characteristics enable the noncoherent detection.

As shown in Fig.~\ref{fig:Cubesplit}, the cube-split constellation has a cell-type structure, which helps detection complexity reduction.
In addition, the bit labeling of cube-split is sophisticated and exhibits high performance in terms of both SER and BER.

\subsubsection{Manopt \cite{gohary2009noncoherent,ngo2020cubesplit}}
A construction method based on optimization on the manifold is termed manopt in this paper.
Through the optimization, there are a number of design metrics for Grassmann constellation.
This paper reviews a method that maximizes MCD between codewords \cite{ngo2020cubesplit}.
Theoretically, it is the best approach in terms of MCD that determines SER.

Let $\cX = \{\X_1, \cdots, \X_{|\cX|}\}$ be a set of Grassmann constellation, where its cardinality $|\cX|$ is an arbitrary parameter determined depending on the transmission rate $R = \log_2 |\cX| / T$.
Then, MCD is defined as
\begin{align}
d_{\mathrm{min}} = \underset{1\leq i < j \leq |\cX|}{\min} d_c(\X_i,\X_j),
\end{align}
where the chordal distance is given by $d_c(\X_i,\X_j) = \left\| \X_i \X_i^{\Hrm} - \X_j \X_j^{\Hrm} \right\|_{\Frm} / \sqrt{2}$.
For $M=1$, the distance is simplified to $d_c(\x_i,\x_j) = \sqrt{1 - |\x_i^{\Hrm} \x_j|^2}$.

The maximization problem for MCD can be formulated by
\begin{equation}
\begin{aligned}
\maximize_{\cX} \quad & \underset{1\leq i < j \leq |\cX|}{\mathrm{min}} d_c(\x_i,\x_j) \\
\textrm{s.t.} \quad & \x_i \in \mathcal{G}(T, M),\;\forall\; i=\{1,\cdots,|\cX|\},
\end{aligned}
\label{opti_cost}
\end{equation}
where the objective function of \eqref{opti_cost} can be approximated by the log-sum-exp function as
\begin{equation}
\begin{aligned}
\minimize_{\cX} \quad & \mathrm{log} \sum_{1\leq i < j \leq |\cX|} \exp\left( - \frac{
    \left\| \X_i \X_i^{\Hrm} - \X_j \X_j^{\Hrm} \right\|_{\Frm}
    }{\epsilon}\right) \\
\textrm{s.t.} \quad & \X_i \in \mathcal{G}(T, M),\;\forall\; i=\{1,\cdots,|\cX|\},
\end{aligned}
\label{opti_lse}
\end{equation}
with $\epsilon$ being a smoothing constant, e.g., $\epsilon=10^{-2}$ and $\epsilon=10^{-3}$.
In particular, for $M=1$, the general optimization problem of \eqref{opti_lse} can be simplified into
\begin{equation}
\begin{aligned}
\minimize_{\cX} \quad &\mathrm{log} \sum_{1\leq i < j \leq |\cX|} \exp\left( \frac{|\x^{\Hrm}_i\x_j|}{\epsilon}\right) \\
\textrm{s.t.} \quad & \x_i \in \mathcal{G}(T, 1),\;\forall\; i=\{1,\cdots,|\cX|\}.
\end{aligned}
\label{opti_lse_M1}
\end{equation}

Both equations \eqref{opti_lse} and \eqref{opti_lse_M1} have the differentiable objective functions and the Grassmann manifold constraints $\x_i \in \mathcal{G}(T, M)$.
Then, these optimization problems can be solved efficiently using optimization tools on a manifold such as manopt \cite{boumal2014manopt}.

The challenge with manopt constellation lies in the bit labeling.
Unlike exp-map and cube-split, the manopt constellation has no structure, making Gray coding difficult.
This task can be formulated as a quadratic assignment problem (QAP) \cite{colman2011quasigray}, classified as NP-hard, and its exhaustive search requires $O(|\cX|!) = O(2^{RT}!)$ trials, which induces a combinatorial explosion.
To alleviate this prohibitive complexity, the fast approximate QAP algorithm \cite{vogelstein2015fast},
the 2-opt algorithm \cite{fishkind2019seeded}, and quasi-Gray labeling method \cite{colman2011quasigray} can be used for manopt.

The advantage of the manopt constellation is the flexibility of the transmission rate.
To adjust the transmit rate, exp-map may require QAM or phase-shift keying (PSK) with different modulation orders within a codeword.
Similarly, cube-split may require different $B_j$.
In both cases, different modulation orders and $B_j$ tend to result in lower MCD, which leads to lower performance.
The manopt constellation circumvents this problem.

\subsection{Noncoherent Detection \cite{warrier2002spectrally,ngo2020cubesplit}}
A point $\X \in \mathbb{C}^{T \times M}$ on the Grassmann manifold is a nonsquare unitary matrix that satisfies $\X^{\Hrm}\X = \mathbf{I}_M$ and $\mathrm{E}[\X \X^{\Hrm}] = M / T \I_T$.
According to the system model of \eqref{eq:sys}, the received symbols are represented as
\begin{align}
    \Y = \S \H + \sigma_v \V
    = \sTM \X \H + \sigma_v \V, \label{eq:sysgrass}
\end{align}
which satisfies the constraint $\mathrm{E}[\|\S\|_{\mathrm{F}}^2] = \mathrm{E}[\|\sqrt{T / M}\X\|_{\mathrm{F}}^2] = T$ and the definition of SNR is the same as \eqref{eq:sys}.

At the receiver, the generalized likelihood ratio test (GLRT) detector \cite{warrier2002spectrally} obtains the estimate of $\X$ as
\begin{align}
    \hat{\X} = \underset{\X \in \cX}{\mathrm{argmax}} \ \mathrm{tr}
    \left[
    \Y\Y^{\Hrm}\X\X^{\Hrm}
    \right],
    \label{eq:GLRT_MI}
\end{align}
where $\mathrm{tr}[\cdot]$ denotes the trace norm.
The detector \eqref{eq:GLRT_MI} is equivalent to the maximum likelihood detection on the Grassmann manifold.
For $M=1$, the detector \eqref{eq:GLRT_MI} can be simplified into \cite{ngo2020cubesplit}
\begin{align}
    \hat{\x} = \underset{\x \in \cX}{\mathrm{argmax}} \ \| \Y^{\Hrm}\x \|^2 \label{eq:GLRT_SI}
\end{align}
Performing noncoherent detection, neither \eqref{eq:GLRT_MI} nor \eqref{eq:GLRT_SI} contain the estimated value of CSI, $\Hh$.
Note that, as clarified in \cite{warrier2002spectrally}, both work with an arbitrary norm of $\| \X \|_{\Frm}^2$.

\subsection{Classic DC-RS \cite{yu2007informationbearing}}
\begin{figure}[tb]
	\centering
	\subfigure[Typical non-data carrying RS.]{
		\includegraphics[clip, scale=0.60]{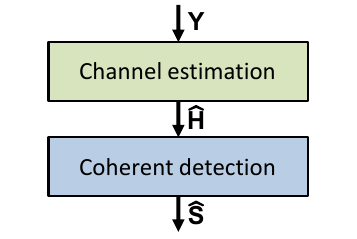}
	}
	\subfigure[DC-RS \cite{yu2007informationbearing}.]{
		\includegraphics[clip, scale=0.60]{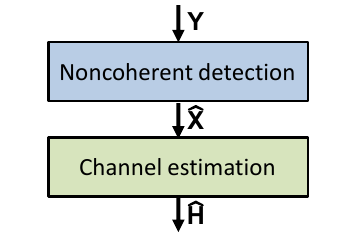}
	}
	\caption{Comparison of estimation procedures.\label{fig:flow}}
\end{figure}
In the conventional training method, the channel matrix $\H$ is estimated by an RS. The estimated $\Hh$ is then used for coherent detection of subsequent high-rate data symbols $\S$ such as 256-QAM and 1024-QAM.
Here, RS can be replaced by DC-RS \cite{yu2007informationbearing}.
Specifically, a Grassmann codeword $\X$ is estimated by the noncoherent detector of \eqref{eq:GLRT_MI}, conveyning additional information that improves SE, and after obtaining $\hat{\X}$ without $\H$, $\H$ is estimated using $\Y$ and $\hat{\X}$ \cite{yu2007informationbearing}.
Each of estimation procedures is summarized in Fig.~\ref{fig:flow}.
Since $\Hat{\X}$ is a tall unitary matrix, if $\Hat{\X} = \X$, both ZF and MMSE equalizers can be considered, and the channel matrix is estimated by\footnote{Note that the inverse operation can be avoided as given in \eqref{eq:R-pinv}.}
\begin{align}
    \Hh = \begin{cases}
        \sqrt{\frac{M}{T}}\hat{\X}^{\Hrm}\Y & \textrm{(ZF)} \\
        \sqrt{\frac{T}{M}} \left(\frac{T}{M}\hat{\X}^{\Hrm} \hat{\X} + \sigma_v^2 \I_M \right)^{-1} \hat{\X}^{\Hrm} \Y & \textrm{(MMSE)}\\
    \end{cases}.
    \label{eq:ZFMMSE}
\end{align}

The main disadvantage of DC-RS, in contrast to the advantage of improving SE, is the increase in computational complexity.
Specifically, for the transmission rate $R = \log_2 |\cX| / T$, the noncoherent detector of \eqref{eq:GLRT_MI} requires $\mathcal{O}(2^{RT} (T^2 N + M^2 T + T^3))$ real-valued multiplications, which indicates that the complexity increases exponentially with $R$ and $T$.
Since the computational complexity for channel estimation, \eqref{eq:ZFMMSE}, is identical to that of the conventional RS, the overall complexity is dominated by the noncoherent detection process.
This can be resolved by low-complexity detectors proposed for the exp-map \cite{kammoun2007noncoherent} and cube-split \cite{ngo2020cubesplit} constellations, which achieve near-optimal performance in some cases.
Then, we assume the use of optimal GLRT detector for the rest of this paper.

\section{Theoretical Analysis of Channel Estimation Accuracy\label{sec:prop}}
In this section, we analyze the channel estimation errors induced by DC-RS on the Grassmann manifold.
As we will see, an MCD-maximizing Grassmann constellation leads to a performance degradation in terms of the channel estimation accuracy.\footnote{To be more specific, this degradation is induced by the uniformity of the optimized symbols, which is depicted in Fig.~\ref{fig:ipro}(c).}
To mitigate this issue, based on our analysis, we then propose a unitary optimization method for the Grassmann constellation. This method improves channel estimation accuracy while maintaining the same MCD as the non-optimized constellation.

In our analysis, we consider the ZF equalization with unknown SNR, defined in \eqref{eq:ZFMMSE}, rather than the MMSE equalization, in order to simplify the analysis.
If SNR is known in advance, the MMSE equalization is a better option, and the performance can be further improved. Our performance comparison is also based on the ZF equalization, ensuring that any performance improvements obtained by DC-RS are not overstated. A more sophisticated channel estimation method can be considered, but the specific estimation method is not expected to affect the relative performance difference between the classic RS and DC-RS.

\subsection{Channel Estimation Accuracy and Its Optimization}
Simultaneous data and channel estimation is enabled in a completely blind manner, and unlike conventional methods such as \cite{shabazpanahi2005blind}, Grassmann-based DC-RS has the advantage of not suffering from phase uncertainties.
Since $\hat{\X}$ is a tall unitary matrix, if $\Hat{\X} = \X$, an original channel estimate $\Hh$ can be obtained thorough the ZF equalization of \eqref{eq:HohatmH} multiplying $\Hat{\X}^{\Hrm}$ from the left side of the received signal $\Y$, i.e.,
\begin{align}
    \Hh
    =\sqrt{\frac{M}{T}}\hat{\X}^{\Hrm}\Y  = \hat{\X}^{\Hrm}\X\H + \Vp = \H + \Vp,
    \label{eq:Hesti}
\end{align}
where $\Vp = \sigma_v \sqrt{M / T}\hat{\X}^{\Hrm}\V$.
Later, we expand $\mathrm{E}[\| \Hh - \H \|_{\Frm}^2]$.
Here, two cases can be considered: $\hat{\X} = \X$ and $\hat{\X} \neq \X$.
First, if the codeword is estimated correctly, i.e., $\hat{\X} = \X$, we have
\begin{align}
    \Hh - \H = \Vp \label{eq:Hh-H-correct}
\end{align}
and $\mathrm{E}[\| \Vp \|_{\Frm}^2] = \sigma_v^2 M^2N / T$.
This event occurs with a probability of $1-\mathrm{SER}$, which can be obtained via Monte Carlo simulations.
Instead, we use the union bound \cite{gohary2009noncoherent}
\begin{align}
    \overline{\mathrm{SER}} = \frac{2}{|\cX|}\sum^{|\cX|-1}_{i=1}\sum^{|\cX|}_{j=i+1}
    p_{i, j},\label{eq:SERbar}
\end{align}
where a pairwise error probability is given by \cite{kammoun2007noncoherent}
\begin{align}
    p_{i, j} = \frac{\sigma_v^{2MN}{2MN-1 \choose MN}}{\mathrm{Re}[\det(\I_M - \X_i^{\Hrm}\X_j\X_j^{\Hrm}\X_i)]^N}
    \label{eq:pij}
\end{align}
for a pair of different codewords $(\X_i, \X_j) \in \cX$ with $1 \leq i < j \leq |\cX|$.
According to Figs.~8 and 9 of \cite{gohary2009noncoherent}, the union bound $\overline{\mathrm{SER}}$ becomes sufficiently tight at SNRs of 15 dB or higher.
In this SNR region, the slight gap between the true $\mathrm{SER}$ and $\overline{\mathrm{SER}}$ can be quantified by a constant scaling factor, $0 < \kappa < 1$, so as to satisfy $\mathrm{SER} = \kappa \cdot \overline{\mathrm{SER}}$.
Now, we have a relationship $1 - \mathrm{SER} = 1 - \kappa \cdot \overline{\mathrm{SER}}$.\footnote{Note that $1 - \kappa \cdot \overline{\mathrm{SER}}$ may become negative at low SNRs, but this case is out of the focus of our analysis since the maximization of MCD used in this paper is equivalent to the minimization of $\overline{\mathrm{SER}}$, which relies on the high SNR assumption.}
Second, if the codeword is estimated incorrectly, the term $\H$ does not disappear unlike \eqref{eq:Hh-H-correct}.
This event occurs with a probability $\kappa \cdot p_{i, j}$, where the original codeword $\X_i$ is wrongly detected as $\X_j$, i.e. $\X_i \neq \X_j$, and we have
\begin{align}
    \Hh - \H = \sqrt{\frac{M}{T}} \X_j^{\Hrm} \left( \sqrt{\frac{T}{M}} \X_i \H + \sigma_v \V \right) - \H
    = \D_{i, j} + \Vp,\label{eq:Hh-H-incorrect}
\end{align}
\begin{align}
    \D_{i, j} = (\X_j^{\Hrm} \X_i - \I_M) \H, \label{eq:Dij}
\end{align}
and $\Vp = \sigma_v \sqrt{M / T}\hat{\X}_j^{\Hrm}\V$.
Since $\H$ and $\V$ are independent, $\D_{i, j}$ and $\Vp$ here are also independent, which indicates that $\mathrm{E}[\| \D_{i,j} + \Vp \|_{\Frm}^2] = \mathrm{E}[\| \D_{i,j} \|_{\Frm}^2] + \mathrm{E}[\| \Vp \|_{\Frm}^2]$ holds asymptotically.
Using these relationships, the expectation of $\| \Hh - \H \|_{\Frm}^2$ can be directly transformed into
\begin{align}
    &\mathrm{E}[\| \Hh - \H \|_{\Frm}^2] =
    (1-\kappa\cdot\overline{\mathrm{SER}}) \mathrm{E}[\| \Vp \|_{\Frm}^2] \nonumber \\
    & ~ ~ ~ ~ ~ ~ ~ ~ ~ ~ ~ ~ ~ ~ ~ ~ ~ ~ ~ ~ + \kappa \cdot \frac{2}{|\cX|}
    \sum^{|\cX|-1}_{i=1}\sum^{|\cX|}_{j=i+1}
    p_{i, j} \mathrm{E}[\| \D_{i,j} + \Vp \|_{\Frm}^2] \nonumber \\
    &= \mathrm{E}[\| \Vp \|_{\Frm}^2]
    - \kappa\cdot\overline{\mathrm{SER}} \cdot \mathrm{E}[\| \Vp \|_{\Frm}^2] + \kappa\cdot\overline{\mathrm{SER}} \cdot \mathrm{E}[\| \Vp \|_{\Frm}^2] \nonumber \\
    &~~~~~~~~~~~~~~\,
    + \kappa\cdot\frac{2}{|\cX|}\sum^{|\cX|-1}_{i=1}\sum^{|\cX|}_{j=i+1}
    p_{i, j} \mathrm{E}[\| \D_{i,j} \|_{\Frm}^2] \nonumber \\
    &= \sigma_v^2 \frac{M^2N}{T} +
    \kappa\cdot\frac{2}{|\cX|}      \sum^{|\cX|-1}_{i=1}\sum^{|\cX|}_{j=i+1}
    p_{i, j} \mathrm{E}[\| \D_{i,j} \|_{\Frm}^2], \label{eq:HohatmH}
\end{align}
leading to a novel optimization metric.

Based on the transformation given in \eqref{eq:HohatmH}, we propose an optimization method to improve the channel estimation accuracy quantified by $\mathrm{E}[\| \Hh - \H \|_{\Frm}^2]$.
The definition of the Grassmann manifold \eqref{eq:GTM} implies that multiplying a unitary matrix from the right side of points on the Grassmann manifold remain invariant \cite{kammoun2007noncoherent}, and it does not change the average transmit power and MCD.
Considering these factors, we design unitary matrices corresponding to all Grassmann codewords in $\cX$ so as to minimize $\mathrm{E}[\| \Hh - \H \|_{\Frm}^2]$ of \eqref{eq:HohatmH}.

Let $|\cX|$ number of $M$-order unitary matrices be design variables of the optimization task, and denote the set as $\{\mathbf{U}_i| 1 \leq i \leq |\cX|, \mathbf{U}_i \in \mathcal{U}(M)\}$.
Here, to minimize $\mathrm{E}[\| \D_{i,j} \|_{\Frm}^2]$ in \eqref{eq:HohatmH}, our optimization is performed for a pair of different codewords $(\X_i, \X_j) \in \cX$ so that $(\X_i \mathbf{U}_i)^{\Hrm} \X_j \mathbf{U}_j = \mathbf{U}_i^{\Hrm} \X_i^{\Hrm} \X_j \mathbf{U}_j$ approaches the identity matrix.
In addition,
the average Frobenius norm $\mathrm{E}[\| \D_{i,j} \|_{\Frm}^2]$ in \eqref{eq:HohatmH} is weighted by $p_{i,j}$.
But, the denominator of \eqref{eq:pij} represents the principle angles of two subspaces \cite{kammoun2007noncoherent}, and unitary matrix multiplications can be ignored in \eqref{eq:pij}.
Overall, the minimization problem of $\mathrm{E}[\| \Hh - \H \|_{\Frm}^2]$ can be newly formulated by
\begin{equation}
\begin{aligned}
    \minimize_{\{\mathbf{U}_1, \cdots, \mathbf{U}_{|\cX|}\}} \quad & \sum^{|\cX|-1}_{i=1}\sum^{|\cX|}_{j = i + 1}
    \frac{
    \left\|
    \mathbf{I}_M -
    \mathbf{U}_i^{\Hrm}
    \X_{i}^{\Hrm}
    \X_{j}\mathbf{U}_j
    \right\|_{\Frm}^2}
    {\mathrm{Re}[\det(\I_M - \X_i^{\Hrm}\X_j\X_j^{\Hrm}\X_i)]}
    \\
    \textrm{s.t.} \quad & \mathbf{U}_i \in \mathcal{U}(M),\;\forall\; i=\{1,\cdots,|\cX|\}.
\end{aligned}
\label{eq:opti_phase}
\end{equation}

As given in \eqref{eq:opti_phase}, the objective function does not include the channel matrix $\H$, and it is free from the expectation operations in \eqref{eq:HohatmH}, indicating that the optimization needs to be performed only once, offline.
As with \eqref{opti_lse} and \eqref{opti_lse_M1},
the optimization task of \eqref{eq:opti_phase}
has a differentiable objective function and has the unitary constraint only, i.e., $\mathbf{U}_i \in \mathcal{U}(M)$, which can be solved efficiently using manifold optimization techniques such as \cite{boumal2014manopt}.
Using the optimized $|\cX|$ unitary matrices, a new Grassmann constellation is designed by $\cX^{\prime} = \left\{\X_i \mathbf{U}_i ~ | ~ 1\leq i \leq L, \ \X_i \in \cX, \ \mathbf{U}_i \in \mathcal{U}(M) \right\}$, which achieves the same MCD as the original constellation $\cX$, resulting in the same SER and achievable rate.
Our design method is applicable to any Grassmann constellation including exp-map \cite{kammoun2007noncoherent} and cube-split \cite{ngo2020cubesplit}.

\subsection{Modeling and Evaluation of Channel Estimation Accuracy}

For the sake of discussion without loss of generality, in this paper, we model the CSI estimation error through a Gauss-Markov uncertainty of \cite{nosrat-makouei2011mimo}
\begin{align}
    \Hb = \sqrt{1 - \beta^2} \H + \beta \E, \label{eq:Hb}
\end{align}
where $\H$ is the true channel matrix and $\E$ is an error matrix whose element follows $\mathcal{CN}(0,1)$.
The parameter $0 \leq \beta \leq 1$ determines the uncertainty of $\Hb$.
That is, $\beta = 0$ corresponds to PCSI and $\beta = 1$ corresponds to no CSI knowledge in advance.
Here, the modeled matrix $\Hb$ can be generated using the estimated channel matrix, $\Hh$ of \eqref{eq:Hesti}, i.e., $\Hb' = \Hh / \alpha$ with $\alpha = \sqrt{\mathrm{E}[ \| \Hh \|_{\Frm}^2] / (N \cdot M)}$.
As with the original channel matrix $\H$, the normalized channel matrices $\Hb$ and $\Hb'$ satisfy $\mathrm{E}[\| \H \|_{\Frm}^2] = \mathrm{E}[\| \Hb \|_{\Frm}^2] = \mathrm{E}[\| \Hb' \|_{\Frm}^2] = NM$, while the Frobenius norm of the estimated channel matrix increases with the noise variance, i.e., $\mathrm{E}[ \| \Hh \|_{\Frm}^2] = NM(1 + \sigma_v^2 M / T)$.
Theoretical analysis becomes complicated if the norm of the channel matrix changes depending on the noise variance.
Then, we use $\H$ and $\Hb$ instead of $\Hh$ for the analysis of achievable rate.

Channel estimation accuracy is quantified by the normalized mean square error (NMSE) of
\begin{align}
    \mathrm{NMSE} = 10 \cdot \mathrm{log}_{10} (\sigma_e^2 ) ~ \mathrm{[dB]}
    \label{eq:NMSE}
\end{align}
and
\begin{align}
    \sigma_e^2 = \frac{\mathrm{E}[\|\Hb-\H\|_{\Frm}^2]}{\mathrm{E}[\|\H\|_{\Frm}^2]} =  \frac{\mathrm{E}[\|\Hb'-\H\|_{\Frm}^2]}{\mathrm{E}[\|\H\|_{\Frm}^2]}.
    \label{eq:sigmae2}
\end{align}
If $\hat{\X} = \X$ holds, we have a lower bound for $\sigma_e^2$, denoted by $\underline{\sigma}_e^2$,
\begin{align}
\sigma_e^2 \geq
\underline{\sigma}_e^2 &=
\left.
\frac{\mathrm{E}[\|\Hb' -\H\|_{\Frm}^2]}{\mathrm{E}[\|\H\|_{\Frm}^2]}
\right|_{\hat{\X} = \X}
\nonumber \\
&= \mathrm{E}\left[ \left\| \left(\frac{1}{\underline{\alpha}} - 1 \right)\H
+ \frac{1}{\underline{\alpha}} \Vp \right\|_{\Frm}^2 \right] / (NM) \nonumber \\
&= \left( \frac{1}{\underline{\alpha}} - 1 \right)^2 + \left( \frac{1}{\underline{\alpha}} \right)^2 \frac{M}{T} \sigma_v^2
\label{eq:sigmae2bar}
\end{align}
with $\underline{\alpha} = \sqrt{1 + \sigma_v^2 M / T} \leq \alpha$
and NMSE is lower bounded by $10 \cdot \log_{10}(\underline{\sigma}_e^2)$.

Substituting \eqref{eq:Hb} into \eqref{eq:sigmae2} yields
\begin{align}
    \sigma_e^2= \frac{\mathrm{E}[\|\Hb-\H\|_{\Frm}^2]}{\mathrm{E}[\|\H\|_{\Frm}^2]} = 2 \cdot (1 - \sqrt{1 - \beta^2}),
\end{align}
which clarifies the relationship between NMSE and the uncertainty of $\Hb$.
That is, depending on a simulated NMSE value, $0 < \sigma_e^2 \leq 2$, the parameter $\beta$ can be determined by
\begin{align}
    \beta = \sqrt{1 - \left( 1 - \frac{\sigma_e^2}{2} \right)^2}.
    \label{eq:beta}
\end{align}
For example, if we use the training-based channel estimation method reviewed in Section~\ref{subsec:train}, which satisfies $\hat{\X}=\X$ in any case, $\beta$ can be calculated by \eqref{eq:sigmae2bar} and \eqref{eq:beta}.
In addition, if we estimate the channel matrix through \eqref{eq:Hesti} using a Grassmann constellation, $\sigma_e^2 = \mathrm{E}[\|\Hb'-\H\|_{\Frm}^2] / \mathrm{E}[\|\H\|_{\Frm}^2]$ is first calculated by Monte Carlo simulations, and then $\beta$ is determined by \eqref{eq:beta}.
Here, $\sigma_e^2$ is affected by $(M, N, T, R)$, $\cX$, and SNR.
Even at low SNRs, $\sigma_e^2$ does not exceed 2 since we normalize the estimated channel matrix $\Hh$ so as to satisfy $\mathrm{E}[\| \Hb' \|_{\Frm}^2] = NM$.

\section{Theoretical Analysis of Achievable Rate\label{sec:ana}}
The achievable rate is an important metric that predicts the minimum SNR of achieving error-free communications in a channel-coded scenario, which forms the basis of typical wireless standards.
In this section, we derive the achievable rate of noncoherent Grassmann constellation for multi-antenna scenarios.
Then, assuming channel estimation errors induced by the estimated Grassmann codewords, we derive the achievable rate of coherent STBC.

Note that the achievable rate is also called constrained capacity, constrained average mutual information, and discrete-input continuous-output memoryless channel capacity, all of which are essentially equivalent.

\subsection{Achievable Rate of Noncoherent Grassmann Constellation for Multi-Antenna Scenarios\label{subsec:ar-mimo}}
Following the pioneering study of Marzetta et al. \cite{marzetta1999Capacity}, it is clarified in \cite{zheng2002communication} that the capacity expression of noncoherent multi-antenna fading channels can be interpreted as sphere packing in the Grassmann manifold.
Using such a noncoherent code, the concept of DC-RS was proposed in \cite{yu2007informationbearing}, and asymptotic SNR loss was analyzed.
These pioneering studies \cite{marzetta1999Capacity,zheng2002communication,yu2007informationbearing} had assumptions of a sufficiently high SNR and the continuous input signaling that corresponds to an infinite constellation size.

As opposed to the studies \cite{marzetta1999Capacity,zheng2002communication,yu2007informationbearing} that clarified the ultimate limit of noncoherent communications, the average mutual information, or achievable rate, assuming discrete inputs and practical SNRs is also important in terms of realistic evaluations.
Such a study is found in \cite{ngo2020cubesplit}, and the achievable rate of noncoherent Grassmann constellation $\cX$ is derived for single-antenna scenarios \cite{ngo2020cubesplit}, leaving multi-antenna scenarios unaddressed.
The achievable rate can be defined by the Monte Carlo integration \cite{ngo2020cubesplit}
\begin{align}
R_g = \frac{B}{T} -
\frac{1}{T |\cX|}
\mathrm{E}_{\H, \V}
\left[
\sum_{i=1}^{|\cX|} \log_2 \left(
\frac{\sum_{j=1}^{|\cX|} p(\Y_i | \X_j)}{p(\Y_i | \X_i)}
\right)
\right],
\label{eq:R}
\end{align}
where we have $\Y_i = \sqrt{T/M} \X_i \H + \sigma_v \V$ determined by random variables $\H$ and $\V$, whose elements follow $\mathcal{CN}(0, 1)$.
In multi-antenna scenarios, the conditional probability $p(\Y | \X)$ in \eqref{eq:R} is given by \cite{gohary2009noncoherent}
\begin{align}
p(\Y|\X) = \frac{\exp(-\tr\left[
\Y^\Hrm (T / M \X\X^\Hrm + \sigma_v^2 \I_T)^{-1}
\Y
\right]
)}{[\pi^T \det (T / M \X\X^\Hrm + \sigma_v^2 \I_T)]^N}.
\label{eq:R-pYX}
\end{align}

Here, the high-complexity inverse matrix calculation in \eqref{eq:R-pYX} may cause calculation errors.
Using the Woodbury matrix identity \cite{petersen2007matrix} and the relationship $\X^\Hrm \X = \I_M$, the inverse calculation in \eqref{eq:R-pYX} can be simplified by
\begin{align}
&\left(\TM \X\X^\Hrm + \sigma_v^2 \I_T \right)^{-1} \nonumber \\
=& \frac{1}{\sigma_v^2} \I_T - \frac{1}{\sigma_v^4}
\TM
\X \left(\I_M + \frac{1}{\sigma_v^2} \TM \X^\Hrm \X\right)^{-1}
\X^\Hrm \nonumber \\
=& \frac{1}{\sigma_v^2} \I_T - \frac{1}{\sigma_v^4}
\TM
\left(1 + \frac{1}{\sigma_v^2} \TM \right)^{-1} \X \X^\Hrm \nonumber \\
=& \frac{1}{\sigma_v^2} \I_T
- \frac{1}{\sigma_v^2} \frac{1}{1 + \sigma_v^2 \MT} \X\X^\Hrm.
\label{eq:R-pinv}
\end{align}

Then, using the inverse matrix \eqref{eq:R-pinv}, the trace norm in \eqref{eq:R-pYX} can be further simplified by
\begin{align}
&-\tr\left[
\Y^\Hrm
(T/M \X\X^\Hrm + \sigma_v^2 \I_T)^{-1}
\Y
\right] \nonumber \\
=&
-\tr\left[
\Y \Y^\Hrm
\left(
\frac{1}{\sigma_v^2} \I_T
- \frac{1}{\sigma_v^2} \frac{1}{1 + \sigma_v^2 \MT} \X\X^\Hrm
\right)
\right] \nonumber \\
=&
-\tr\left[
\frac{\Y \Y^\Hrm}{\sigma_v^2}
- \frac{\Y \Y^\Hrm \X \X^\Hrm}{\sigma_v^2 \left( 1 + \sigma_v^2 \MT \right)}
\right] \nonumber \\
=&- \frac{\|\Y \|_\Frm^2}{\sigma_v^2}
+ \frac{\| \Y^\Hrm \X \|_\Frm^2}{\sigma_v^2 \left(1 + \sigma_v^2 \MT\right)}.
\end{align}

Using the Sylvester's determinant theorem, the determinant in the denominator of \eqref{eq:R-pYX} can also be simplified to
\begin{align}
&\det \left(\TM \X\X^\Hrm + \sigma_v^2 \I_T \right)
= \det(\sigma_v^2 \I_T)
\det \left(
\I_T + \frac{\X \X^\Hrm}{\sigma_v^2 M / T}
\right) \nonumber \\
=& \sigma_v^{2T}
\det \left(
\I_M + \frac{\X^\Hrm \X}{\sigma_v^2 M / T}
\right)
= \sigma_v^{2T}
\left(1 + \frac{1}{\sigma_v^2 M / T}\right)^M,
\end{align}
which now becomes irrelevant to $\X$.
Finally, we obtain new expressions
\begin{align}
p(\Y|\X) = \frac{\exp(
- \|\Y \|_\Frm^2 / \sigma_v^2
+ \| \Y^\Hrm \X \|_\Frm^2 / \sigma_v^2 / \left(1 + \sigma_v^2 M / T \right)
)}{\pi^{TN}
\sigma_v^{2TN}
\left(1 + 1 / \sigma_v^2 / M \cdot T \right)^{MN}
},\label{eq:nar-pYX}
\end{align}
$p(\Y_i | \X_j) / p(\Y_i | \X_i) = \exp(\eta_{i,j})$, and\footnote{Note that $\eta_{i,i} = 0$ holds for any $i$ while $\eta_{i,j} = -\eta_{j,i}$ does not hold for $i \neq j$. We need to calculate both $\eta_{i,j}$ and $\eta_{j,i}$.}
\begin{align}
\eta_{i,j} =
\frac{\| \Y_i^\Hrm \X_j \|_\Frm^2 - \| \Y_i^\Hrm \X_i \|_\Frm^2}{\sigma_v^2 \left(1 + \sigma_v^2 M/ T \right)}.
\label{eq:R-eta}
\end{align}

Using $\eta_{i,j}$ of \eqref{eq:R-eta}, the achievable rate can be rewritten as
\begin{align}
R_g &= \frac{B}{T} -
\frac{1}{T |\cX|}
\mathrm{E}_{\H, \V}
\left[
\sum_{i=1}^{|\cX|} \log_2
\sum_{j=1}^{|\cX|} \exp (\eta_{i,j})
\right] \\
&= \frac{B}{T} -
\frac{1}{T |\cX| \mathrm{ln}(2)}
\mathrm{E}_{\H, \V}
\left[
\sum_{i=1}^{|\cX|}
\mathrm{lse}(\eta_{i,1} , \cdots, \eta_{i,|\cX|})
\right],\label{eq:Rg-lse}
\end{align}
where $\mathrm{ln(\cdot)}$ and $\mathrm{lse(\cdot)}$ are natural logarithm and log-sum-exp functions, respectively.
The log-sum-exp function is crucial for suppressing calculation errors.

\subsection{Achievable Rate of Coherent STBC Assuming Channel Estimation Errors\label{subsec:ar-error}}
Ideally, the Shannon capacity, or the channel capacity, increases monotonically in logarithmic order with respect to SNR.
But, it is known that estimation errors of CSI have a significant negative impact on the capacity, causing it to peak at a specific constant value despite increasing SNR \cite{yoo2006capacity}.
It can be inferred that the same happens for achievable rate analysis assuming CSI errors.

Conventionally, for a set of discrete space-time codewords, $\S_i \in \cS$, the achievable rate assuming PCSI is calculated as \cite{ng2006mimo}
\begin{align}
R = \frac{B}{T} -
\frac{1}{T |\cS|}
\mathrm{E}_{\H, \V}
\left[
\sum_{i=1}^{|\cS|} \log_2 \left(
\frac{\sum_{j=1}^{|\cS|} p(\Y_i | \S_j, \H)}{p(\Y_i | \S_i, \H)}
\right)
\right],
\label{eq:Rs}
\end{align}
where we have \cite{ng2006mimo}
\begin{align}
    p(\Y|\S, \H) = \frac{1}{(\pi\sigma_v^2)^{NT}}\exp \left( - \frac{\|\Y-\S \H\|_{\Frm}^2}{\sigma_v^2}  \right)
\end{align}
and \cite{ng2006mimo}
\begin{align}
\frac{p(\Y_i | \S_j, \H)}{p(\Y_i | \S_i, \H)} =
\exp \left(\frac{-\|\Y_i-\S_j \H \|_{\Frm}^2 +
\|\Y_i - \S_i \H \|_{\Frm}^2}{\sigma_v^2}  \right).
\label{eq:Rs-pratio}
\end{align}
Here, the relationship $\Y_i = \S_i \H + \sigma_v \V$ is assumed, and the likelihood ratio of \eqref{eq:Rs-pratio} can be further simplified.

The case of CSI errors is tricky and decoding is performed on the received signal $\Y_i = \S_i \H + \sigma_v \V$ using the estimated channel matrix $\Hb$ containing errors $\E$.
The effective noise variance is different from the PCSI case, and more random variables are involved.
The achievable rate assuming CSI errors can be newly expressed by
\begin{align}
R_e = \frac{B}{T} -
\frac{1}{T |\cS|}
\mathrm{E}_{\Hb, \V}
\left[
\sum_{i=1}^{|\cS|} \log_2 \left(
\frac{\sum_{j=1}^{|\cS|} p(\Y_i | \S_j, \Hb)}{p(\Y_i | \S_i, \Hb)}
\right)
\right],
\label{eq:Re}
\end{align}
where we have
\begin{align}
    p(\Y|\S, \Hb) = \frac{1}{(\pi(\sigma_v^2 + \sigma_e^2))^{NT}}\exp \left( - \frac{\|\Y-\S \Hb\|_{\Frm}^2}{\sigma_v^2 + \sigma_e^2}  \right)
\end{align}
and
\begin{align}
\frac{p(\Y_i | \S_j, \Hb)}{p(\Y_i | \S_i, \Hb)} =
\exp \left(\frac{- z_{i,j} +
z_{i,i} }{\sigma_v^2 + \sigma_e^2}  \right).
\label{eq:pYiSjpYiSi}
\end{align}
In \eqref{eq:pYiSjpYiSi}, the random variable $z_{i,j}$ is defined as
\begin{align}
z_{i,j} &=
\|\Y_i-\S_j \Hb \|_{\Frm}^2 \nonumber \\
&= \| ( \S_i - \sqrt{1-\beta^2} \S_j ) \H + \sigma_v \V - \beta \S_j \E \|_{\Frm}^2.
\end{align}
Since $\V$ and $\E$ are uncorrelated random Gaussian variables, and
we have the power constraint
$\mathrm{E}[\|\S_j\|_{\mathrm{F}}^2] = T$, $z_{i,j}$ can be simplified into
\begin{align}
z_{i,j} \simeq \| ( \S_i - \sqrt{1-\beta^2} \S_j ) \H + \sqrt{\sigma_v^2 + \beta^2} \V \|_{\Frm}^2,\label{eq:zij-approx}
\end{align}
where the negative effects of $\sigma_v \V$ and $\beta \S_j \E$ are expressed by a single noise component $\sqrt{\sigma_v^2 + \beta^2} \V$.
The generation of additional noise $\E$ is no longer needed through the Monte Carlo integration.

\begin{figure}[tb]
	\centering
    \includegraphics[clip, scale=0.60]{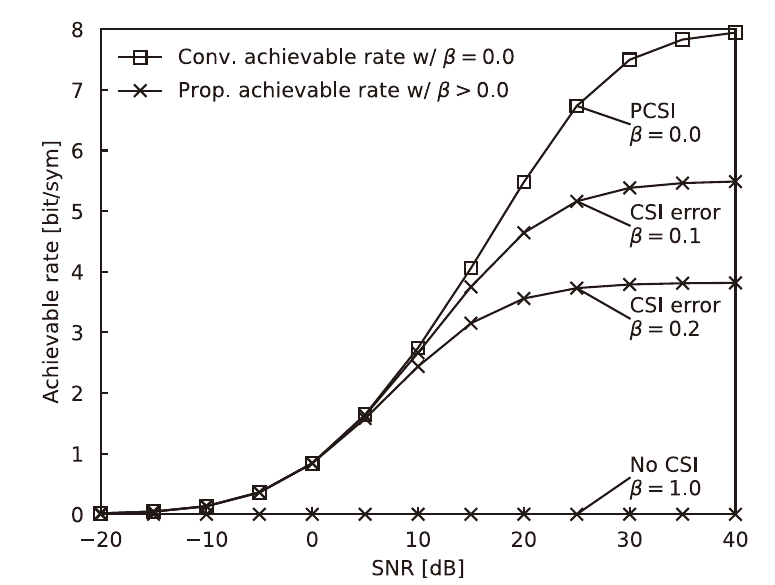}
    \caption{Example of achievable rate comparisons where we considered 256-QAM and $M=N=1$. \label{fig:ar-ex}}
\end{figure}
As an illustrative example, Fig.~\ref{fig:ar-ex} shows achievable rates of coherent 256-QAM signaling assuming PCSI \eqref{eq:Rs} and CSI errors \eqref{eq:Re}.
As shown in Fig.~\ref{fig:ar-ex},
when $\Hb$ has maximum uncertainty $\beta=1$, no information is conveyed at all.
Compared to the PCSI case of $\beta=0$, the CSI error of $\beta = 0.2$ reduces the achievable rate by half.
Thus, minimizing NMSE is critical in terms of maximizing SE.

\section{Performance Results\label{sec:performance}}
In this section, we compare the training-based and Grassmann-based channel estimation methods in terms of the channel estimation accuracy, which is quantified by NMSE of \eqref{eq:NMSE}, and SE.
Here, we calculate the achievable rates of noncoherent Grassmann constellation and coherent signaling assuming CSI errors, both of which are newly derived in Section~\ref{sec:ana}.
While a decrease in channel estimation accuracy is expected with the estimation of Grassmann codeword, we verify whether a practical gain in terms of SE can be obtained, based on the derived achievable rates.

\begin{figure}[tb]
	\centering
	\includegraphics[clip, scale=0.60]{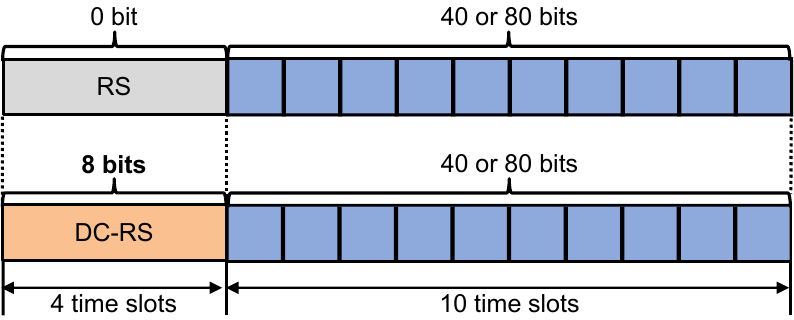}
	\caption{Example of slot structure considered in our comparisons.\label{fig:slot}}
\end{figure}
In our comparisons, we consider $M=N=1$ or $M=N=2$ antennas\footnote{More receive antennas improve SE, but its specific number does not affect the relative performance differences.}, and a slot composed of a $T=4$ RS and subsequent ten data symbols, comprising a total of 14 time slots.
This setup corresponds to a scenario where the first $T$ symbols are assigned to reference signaling and the following symbols are used for data transmission.
Fig.~\ref{fig:slot} shows a slot structure considered in our comparisons.
In the classic training method, an RS with symbol length four consists of random QPSK symbols, which are perfectly known at both transmitter and receiver.
In our proposed method, the DC-RS consists of an exp-map, a cube-split, or a manopt Grassmann constellation, and it carries additional $B=8$ bits through $T=4$ symbol time.
After obtaining an estimate of channel matrix, $M$ number of 16-QAM symbols are transmitted through $M$ antennas at each of 10 time slots.
That is, $10 \cdot 4$ bits are conveyed for $M=1$, while $10 \cdot 4 \cdot 2$ bits are conveyed for $M=2$, which corresponds to the classic spatial multiplexing transmission.

\begin{figure}[tb]
	\centering
	\subfigure[$(M,T)=(1,4)$.]{
		\includegraphics[clip, scale=\HISTW]{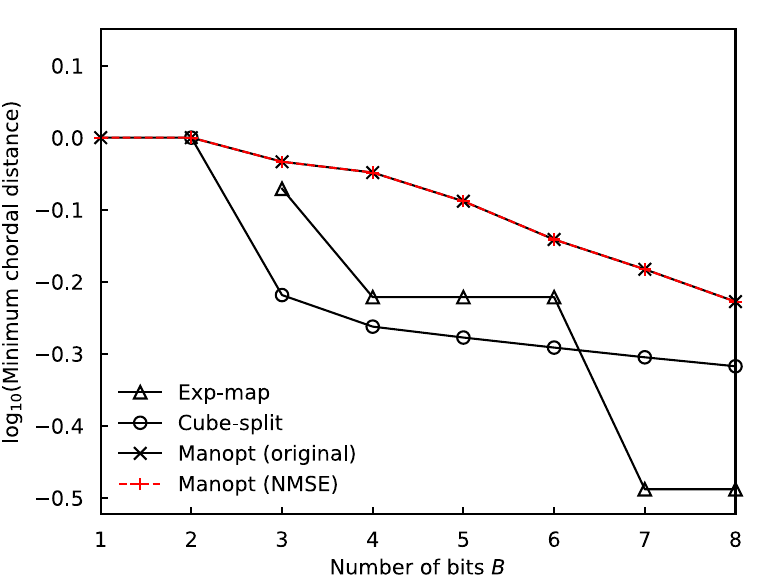}
	}
	\subfigure[$(M,T)=(2,4)$.]{
		\includegraphics[clip, scale=\HISTW]{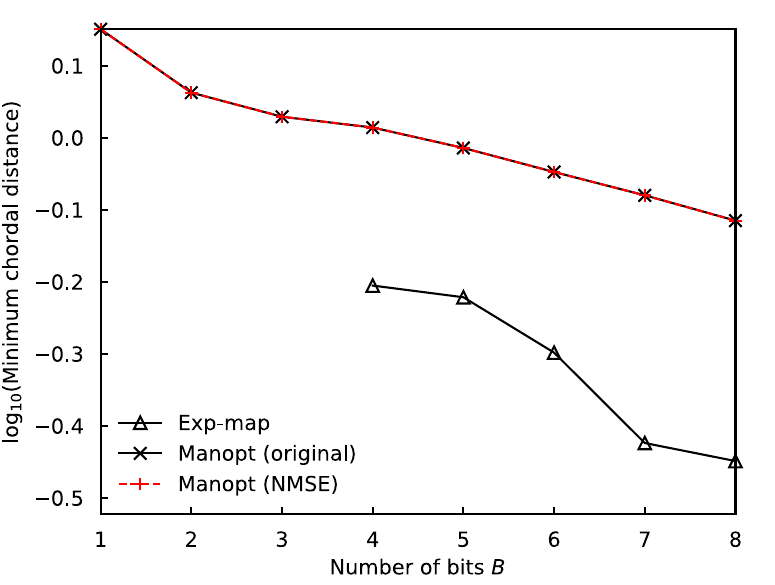}
	}
	\caption{MCD comparisons upon increasing the number of bits. \label{fig:ratemcd}}
\end{figure}
\begin{figure}[tb]
	\centering
	\includegraphics[clip, scale=0.60]{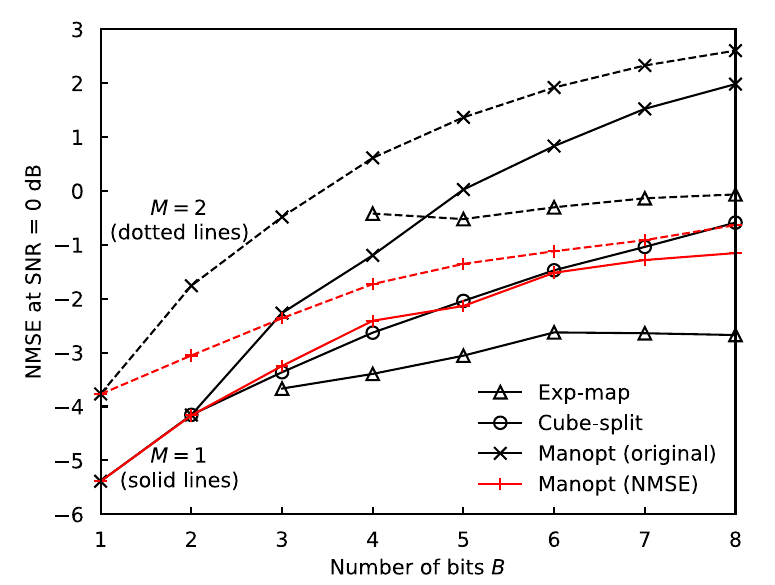}
	\caption{NMSE comparisons upon increasing the number of bits where we considered $M=N=1$ and $M=N=2$. \label{fig:ratenmse}}
\end{figure}

First, in Figs.~\ref{fig:ratemcd} and \ref{fig:ratenmse}, we investigated the effects of the NMSE-minimizing method proposed in Section~\ref{sec:prop}, where we varied the number of antennas and increased the number of bits per codeword from $B=1$ to $B=8$, while the number of time slots was fixed at $T=4$.
In Fig.~\ref{fig:ratemcd}, we compared the exp-map, cube-split, original manopt, and NMSE-minimizing manopt constellations in terms of MCD.
In the original study of Kammoun et al., the exp-map constellation is intended to use QAM symbols only \cite{kammoun2003new,kammoun2007noncoherent}.
But, we also considered the exp-map constellation with BPSK, 8-PSK, and 32-PSK symbols, which achieve competitive MCDs, to enable comparisons with other construction methods.\footnote{To be more specific, in the $(M,T,B)=(2,4,4)$ case, the matrix $\mathbf{C}$ of \eqref{eq:exp-map-c} should contain four BPSK symbols, but its MCD becomes worse than the $B=5$ case. Then, we used $s_1$ and $s_3$ drawn from QPSK symbols and set $s_2 = s_4 = 0$.}
In addition, because the construction of cube-split for $M=2$ has been left to be a future work as mentioned in \cite{ngo2020cubesplit}, we omitted it in Fig.~\ref{fig:ratemcd}(b).
As shown in Fig.~\ref{fig:ratemcd}, the proposed NMSE-minimizing method achieved the best MCD at all $B$ for $M=1$ and $M=2$, which was perfectly identical to the original manopt constellation.
This indicates that no performance penalty is induced by the proposed NMSE-minimizing method.
In addition, in Fig.~\ref{fig:ratemcd}(a), it was found that cube-split outperformed exp-map at $B=7$ and $8$, while exp-map was better at other $B$.
The reason why the MCD of exp-map is fluctuating with respect to $B$ is because PSK symbols were used.

Similarly, Fig.~\ref{fig:ratenmse} shows NMSE comparisons of the same setup as Fig.~\ref{fig:ratemcd}, where we additionally considered $\mathrm{SNR} = 0$ dB.
The definition of NMSE is given in \eqref{eq:NMSE}.
As shown in Fig.~\ref{fig:ratenmse}, for $M=1$, the exp-map constellation exhibited the best NMSE, while the NMSEs of the proposed manopt and conventional cube-split were almost identical.
For $M=2$, the proposed manopt achieved the best NMSE at all $B$, outperforming the exp-map constellation.
Compared to the original manopt, the proposed optimization significantly improved NMSE, with a gain of approximately 3 dB observed at $B=8$.\footnote{The proposed optimization method has been applied to the exp-map and cube-split constellations as well, but the improvement was very limited, indicating that exp-map and cube-split were already efficient in terms of NMSE.}
In the following, we mainly focus on the maximum case, $B=8$, to boost SE.

\begin{figure*}[tb]
	\centering
	\subfigure[Exp-map.]{
		\includegraphics[clip, scale=\HISTW]{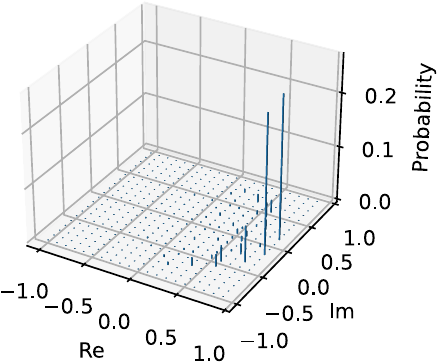}
	}
	\subfigure[Cube-split.]{
		\includegraphics[clip, scale=\HISTW]{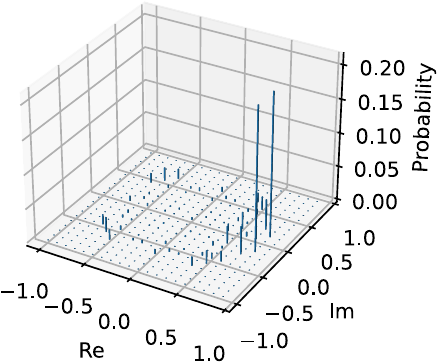}
	}
    \subfigure[Original manopt.]{
		\includegraphics[clip, scale=\HISTW]{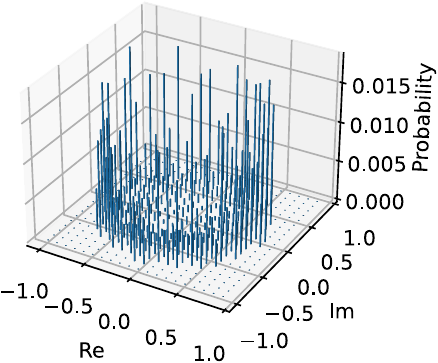}
	}
    \subfigure[NMSE-minimizing manopt.]{
		\includegraphics[clip, scale=\HISTW]{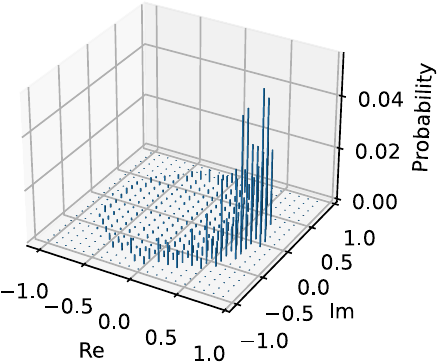}
	}
	\caption{Inner products distribution of Grassmann constellation with $(M, T, B) = (1, 4, 8)$ and $\mathrm{SNR}=15$ dB.\label{fig:ipro}}
\end{figure*}
Why was a large gain of 3 dB obtained in Fig.~\ref{fig:ratenmse}?
To confirm the evidence of this large gain, we calculated 3D histograms of the inner products $\hat{\x}^{\Hrm} \x$ in Fig.~\ref{fig:ipro} when detection errors occurred, i.e., $\hat{\x} \neq \x$.
For a simple visualization, we considered $(M, T, B) = (1, 4, 8)$ and $\mathrm{SNR}=15$ dB.
As analyzed in Section~\ref{sec:prop}, smaller $|\hat{\x}^{\Hrm} \x - 1|^2$ values improve NMSE.
As shown in Fig.~\ref{fig:ipro}, for the exp-map and cube-split constellations, $\hat{\x}^{\Hrm} \x$ was biased around 1 when errors occurred, which contributed to lower NMSE.
By contrast, the original manopt constellation had many $\hat{\x}^{\Hrm} \x$ over the unit circle, which were almost uniformly distributed and far from 1.
This uniformity was induced by the optimization of MCD, which was started from random symbols uniformly distributed on the Grassmann manifold.
Here, we applied the optimization method of Section~\ref{sec:prop} for the original manopt constellation, and it succeeded in biasing $\hat{\x}^{\Hrm} \x$ more toward the area around 1.
Thus, the large gain was observed in Fig.~\ref{fig:ratenmse}.

\begin{figure}[tb]
	\centering
	\subfigure[Number of transmit antennas $M=1$.]{
		\includegraphics[clip, scale=0.60]{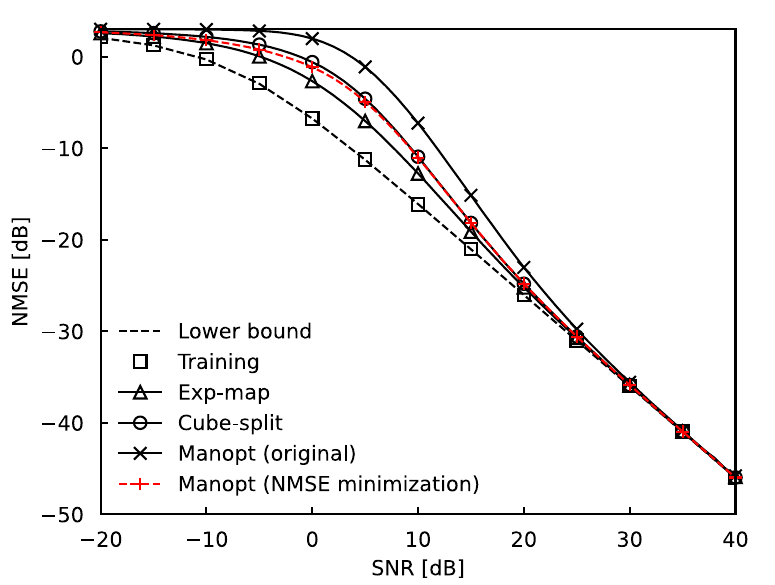}
	}
	\subfigure[Number of transmit antennas $M=2$.]{
		\includegraphics[clip, scale=0.60]{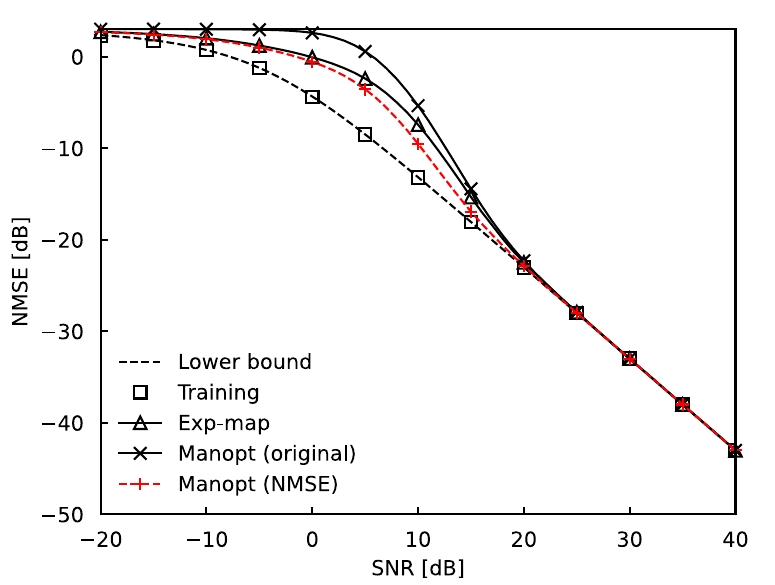}
	}
	\caption{NMSE comparisons where we had $(T,B)=(4,8)$. \label{fig:nmse}}
\end{figure}
Next, we compared NMSE of the training-based and the proposed Grassmann-based channel estimation methods in Fig.~\ref{fig:nmse}, where we varied SNR from $-20$ to $40$ dB.
NMSE values were simulated every 1 dB with $10^6$ trials, but markers were shown at 5 dB intervals.
The NMSE lower bound \eqref{eq:sigmae2bar} was plotted for reference.
As shown in Fig.~\ref{fig:nmse}, the NMSE of the training method was exactly the same as the theoretical lower bound, as anticipated in the analysis of Section~\ref{sec:prop}.
The NMSE of the original manopt was significantly worse than the other construction methods because of the uniformity of the inner product distribution, as depicted in Fig.~\ref{fig:ipro}(c).
By contrast, after the proposed NMSE-minimizing method was applied, the resultant NMSE was greatly improved particularly in the middle SNR region.

The achievable rate $R_e$ of \eqref{eq:Re} assuming CSI errors relies on the channel uncertainty parameter $\beta$, where $\beta$ relies on $\sigma_e^2$, which is used to calculate NMSE as in \eqref{eq:NMSE}.
In the following, the achievable rate $R_e$ was calculated using the $\sigma_e^2$ values obtained from Figs.~\ref{fig:nmse}(a) and (b).
Additionally, since a state-of-the-art channel estimation method would achieve the performance close to the PCSI case, i.e., $\beta=0$ and $\sigma_e^2=0$, we also considered it in our comparisons.

\begin{figure*}[tbp]
	\centering
 \subcapcentertrue
	\subfigure[Achievable rate $R_{g}$ of noncoherent Grassmann constellation derived in \eqref{eq:Rg-lse}.]{
		\includegraphics[clip, scale=\ARW]{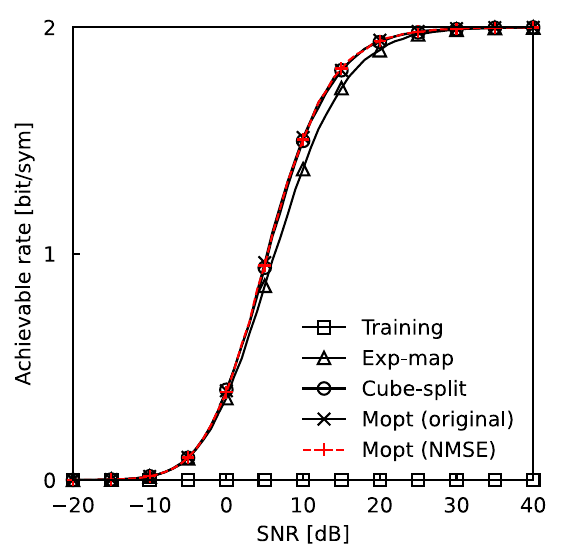}
	}
	\subfigure[Achievable rate $R_{e}$ of coherent 16-QAM with CSI errors derived in \eqref{eq:Re}.]{
		\includegraphics[clip, scale=\ARW]{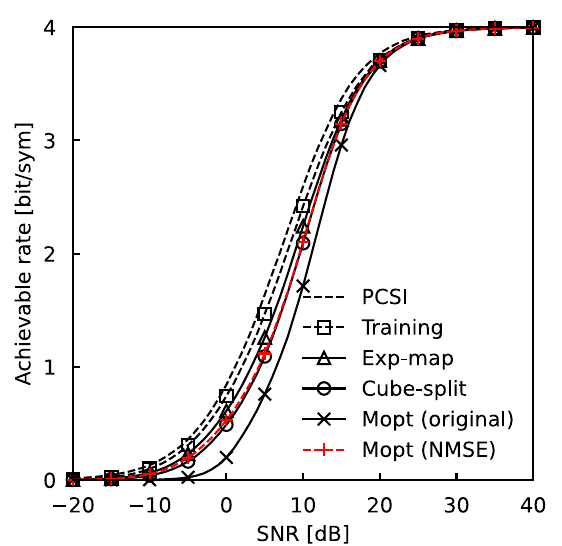}
	}
    \subfigure[Total achievable rate $4 \cdot R_{g} + 10 \cdot R_{e}$ based on the slot structure of Fig.~\ref{fig:slot}.]{
		\includegraphics[clip, scale=\ARW]{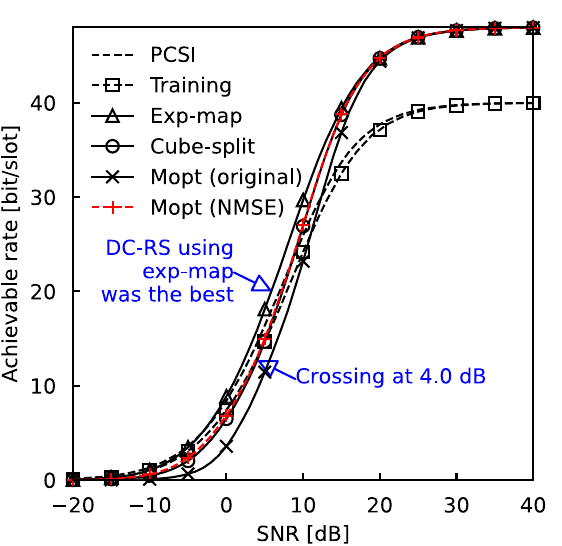}
	}
	\caption{Achievable rate comparisons for $M=1$.\label{fig:ar-M1}}
\end{figure*}
Fig.~\ref{fig:ar-M1} shows comparisons of the achievable rates for $M=1$.
Specifically, the achievable rate $R_{g}$ of noncoherent Grassmann constellation and the achievable rate $R_{e}$ of coherent 16-QAM assuming CSI errors were calculated based on \eqref{eq:Rg-lse} and \eqref{eq:Re}, and the total achievable rate $4 \cdot R_{g} + 10 \cdot R_{e}$ was then calculated according to the slot structure given in Fig.~\ref{fig:slot}.
As shown in Fig.~\ref{fig:ar-M1}(a), the original manopt (indicated with `Mopt') and the proposed NMSE-minimizing manopt achieved the same achievable rate, and we confirmed that the NMSE optimization can be completed without any performance penalty.
In addition, the manopt and cube-split constellations achieved the best achievable rate, and the exp-map constellation exhibited suboptimal rates.
By contrast, interestingly, the coherent 16-QAM relying on the channel estimates of exp-map outperformed these of cube-split and manopt constellations, due to the small NMSE, as shown in Fig.~\ref{fig:ar-M1}(b).
Accordingly, in the total rate comparison of Fig.~\ref{fig:ar-M1}(b), the DC-RS using the exp-map constellation even outperformed the 16-QAM signaling assuming PCSI in the whole SNR region.
The DC-RS using cube-split and manopt outperformed the training method at SNRs greater than $4.0$ dB.
That is, in the $M=1$ case, the DC-RS using exp-map has the potential for outperforming the classic training method and boosting SE at every SNR.

\begin{figure}[tbp]
	\centering
	\includegraphics[clip, scale=0.60]{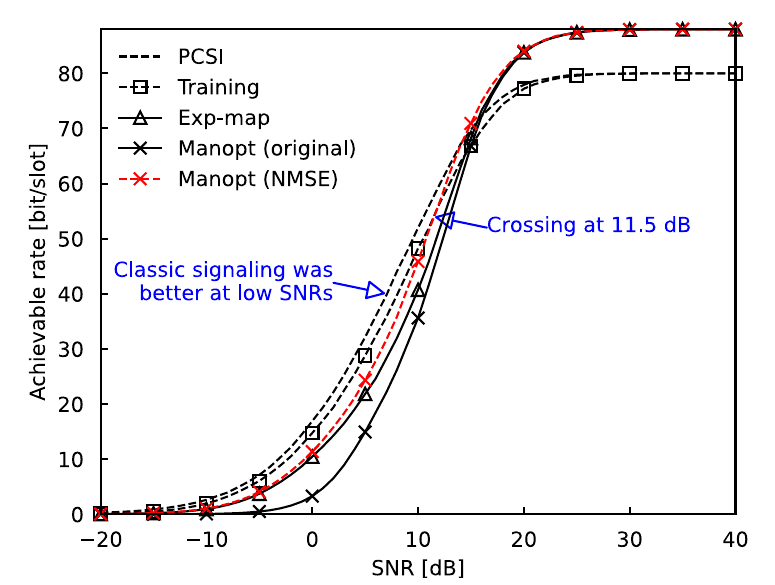}
	\caption{Total achievable rate for $M=2$.\label{fig:ar-M2}}
\end{figure}

Finally, in Fig.~\ref{fig:ar-M2}, we compared the total achievable rate for $M=2$.
As shown, the proposed manopt outperformed the exp-map case, as can be expected from Fig.~\ref{fig:nmse}(b).
However, compared to the $M=1$ case in Fig.~\ref{fig:ar-M1}(c), the classic training method exhibited competitive performance at low SNRs, and the proposed manopt outperformed it at SNRs greater than $11.5$ dB.
That is, for $M=2$, the classic training method is effective at low SNRs, and the proposed DC-RS becomes effective at high SNRs, indicating potential need for adaptive switching.

\section{Conclusions\label{sec:conc}}
In this paper, we considered to use the concept of DC-RS on the Grassmann manifold for simultaneous channel and data estimation. To improve the channel estimation accuracy, we directly expanded its Frobenius norm expression and newly formulated the task as an optimization problem on the manifold. This problem formulation accelerates convergence and improves NMSE without any performance penalty. We also derived achievable rates of noncoherent Grassmann constellation for $M \geq 2$ and coherent STBC assuming CSI errors. These derivations allow comparison of SE when non-data carrying pilots are replaced by DC-RS, suggesting the potential for boosting SE.

\footnotesize{
	\bibliographystyle{IEEEtranURLandMonthDiactivated}
	\bibliography{main}
}

\begin{IEEEbiography}[{\includegraphics[width=1in,height=1.25in,clip,keepaspectratio]{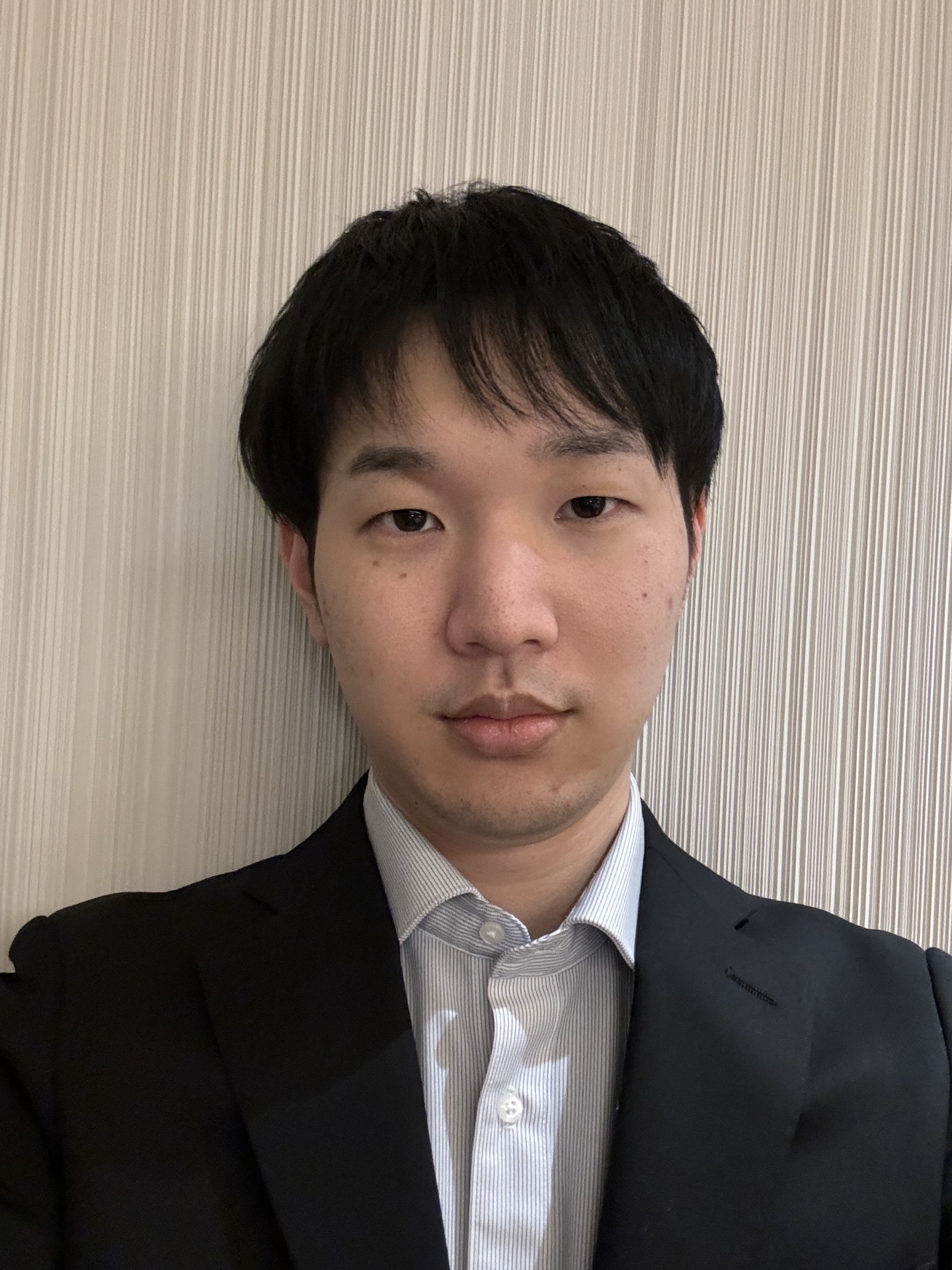}}]{Naoki Endo}
received the B.E. and M.E. degrees from Yokohama National University, Kanagawa, Japan, in 2021 and 2023, respectively. His research interests include MIMO and noncoherent codes on the Grassmann manifold.
\end{IEEEbiography}

\begin{IEEEbiography}[{\includegraphics[width=1in,height=1.25in,clip,keepaspectratio]{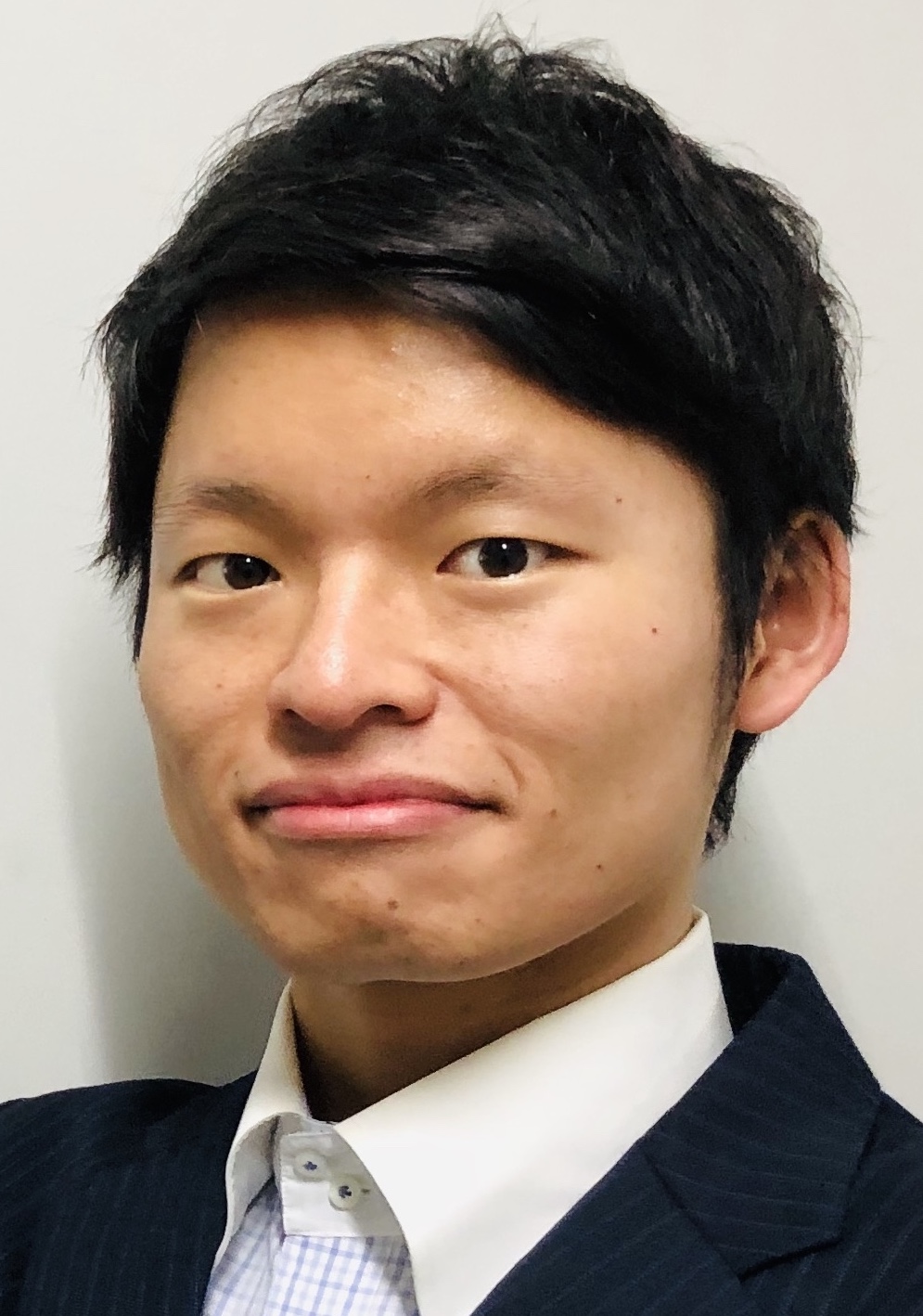}}]{Hiroki Iimori} (S'18--M'22) received his Ph.D. with special distinction from Jacobs University Bremen, Bremen, Germany, in 2022, after receiving his B.Eng. and M.Eng. degrees (Hons.) in electrical and electronic engineering from Ritsumeikan University, Kyoto, Japan, in 2017 and 2019. In 2020, he was a visiting scholar at the Department of Electrical and Computer Engineering, University of Toronto, Toronto, ON, Canada. In 2021, he was a research intern at the Ericsson Radio S\&R Research Laboratory, Yokohama, Japan, where he currently holds the position of Experienced Researcher. His research interests include optimization theory, wireless communications, and signal processing. He was awarded the YKK Doctoral Fellowship by the Yoshida Scholarship Foundation, the IEICE Young Researcher of the Year Award by the IEICE Smart Radio Committee in 2020, among others.
\end{IEEEbiography}

\begin{IEEEbiography}[{\includegraphics[width=1in,height=1.25in,clip,keepaspectratio]{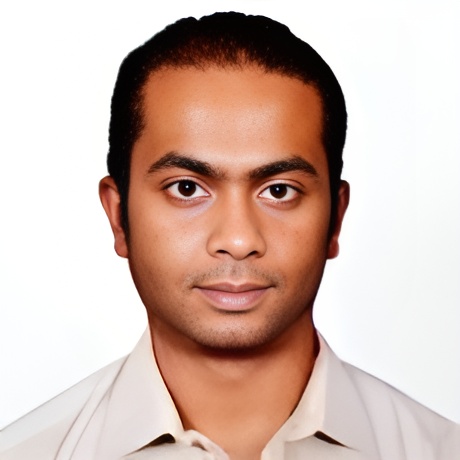}}]{Chandan Pradhan} received the B.Tech. degree from IIIT Bhubaneswar, India, in 2013, and the M.S. degree from IIIT Hyderabad, India, in 2016, and the Ph.D. degree form The University of Sydney, Sydney, NSW, Australia, in 2021. He has worked as Research Engineer in CEWiT, India, in 2016. He is currently working as Senior Researcher in Ericsson research, Japan. His research interests include beamforming techniques in MIMO systems and application of deep-learning in wireless networks.
\end{IEEEbiography}

\begin{IEEEbiography}[{\includegraphics[width=1in,height=1.25in,clip,keepaspectratio]{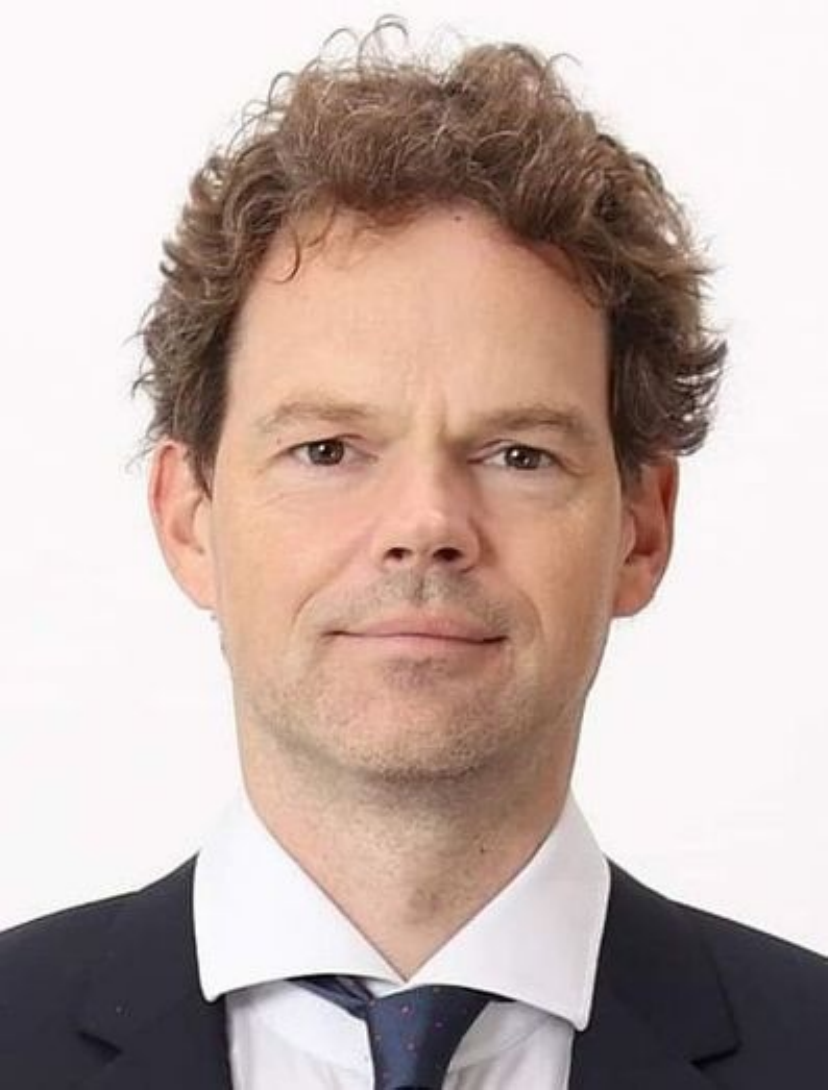}}]{Szabolcs Malomsoky} is the head of Ericsson Research Japan. Earlier he was leading research units in Hungary, Sweden and the US. He received a Ph.D. degree from the Budapest University of Technology and Economics in 2003. Szabolcs worked with strategy setting and technical leadership in research areas including wireless technologies, network analytics, cloud computing, network management and programmable networks.
\end{IEEEbiography}

\begin{IEEEbiography}[{\includegraphics[width=1in,height=1.5in,clip,keepaspectratio]{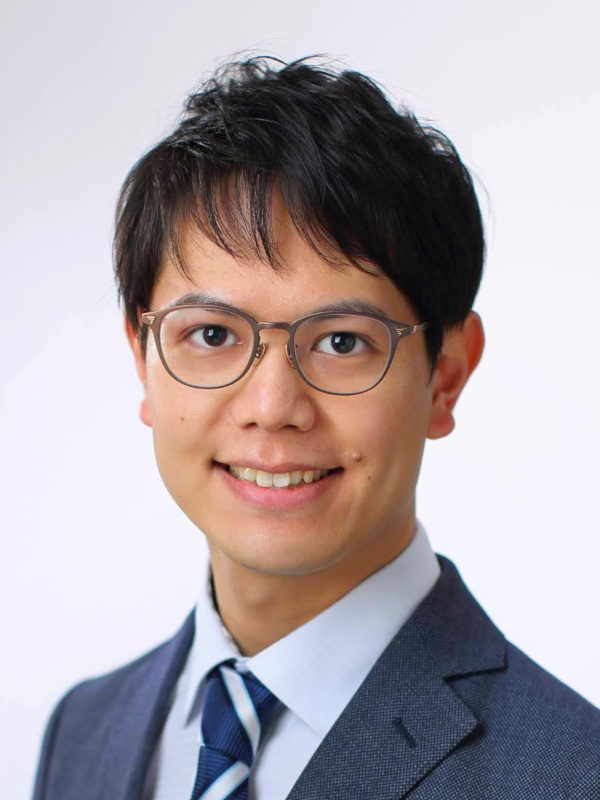}}]{Naoki~Ishikawa}
(S'13--M'17--SM'22) is an Associate Professor with the Faculty of Engineering, Yokohama National University, Kanagawa, Japan. He received the B.E., M.E., and Ph.D. degrees from the Tokyo University of Agriculture and Technology, Tokyo, Japan, in 2014, 2015, and 2017, respectively. In 2015, he was an academic visitor with the School of Electronics and Computer Science, University of Southampton, UK. From 2016 to 2017, he was a research fellow of the Japan Society for the Promotion of Science. From 2017 to 2020, he was an assistant professor in the Graduate School of Information Sciences, Hiroshima City University, Japan. He was certified as an Exemplary Reviewer of \textsc{IEEE Transactions on Communications} in 2017 and 2021. His research interests include quantum algorithms and wireless communications.
\end{IEEEbiography}

\end{document}